\documentclass[%
 reprint,
superscriptaddress,
 amsmath,amssymb,
 aps,
twocolumn,
]{revtex4-2}
\usepackage{bm,amsfonts,amsmath,color,graphicx}
\usepackage{comment,xcolor}

\usepackage{float}
\makeatletter
\let\newfloat\newfloat@ltx
\makeatother
\usepackage{algorithm}
\usepackage{algpseudocode}

\newcommand{\T}[0]{{{\bm\Theta}}}

\newcommand{\X}[0]{{\bf X}}
\newcommand{\tX}[0]{\tilde{\bf X}}
\newcommand{\C}[0]{{\bf C}}
\newcommand{\F}[0]{{\bf F}}

\newcommand{\D}[0]{{\bf D}}

\newcommand{\oU}[0]{{U}}

\begin{document}
\title{Exploring parameter dependence of atomic minima with implicit differentiation}

\author{Ivan Maliyov}
\email{ivan.maliyov@cnrs.fr}
\author{Petr Grigorev}
\author{Thomas D Swinburne}
\email{thomas.swinburne@cnrs.fr}
\affiliation{Aix-Marseille Université, CNRS, CINaM UMR 7325, Campus de Luminy, Marseille 13288, France}
%date{\today}
\begin{abstract}
Interatomic potentials are essential to go beyond \textit{ab initio} size limitations, 
but simulation results depend sensitively on potential parameters. Forward propagation of parameter variation is key for uncertainty quantification, whilst backpropagation has found application for emerging inverse problems such as fine-tuning or targeted design. 
Here, the implicit derivative of functions defined as a fixed point 
is used to Taylor-expand the energy and structure of atomic minima in potential parameters, evaluating terms via 
automatic differentiation, dense linear algebra or a sparse operator approach.
The latter allows efficient forward and backpropagation through relaxed structures of arbitrarily large systems. 
The implicit expansion accurately predicts lattice distortion and defect formation energies and volumes with classical and machine-learning potentials, enabling high-dimensional uncertainty propagation without prohibitive overhead. 
We then show how the implicit derivative can be used to solve challenging inverse problems, minimizing an implicit loss to fine-tune potentials and stabilize solute-induced structural rearrangements at dislocations in tungsten.
\end{abstract}

\maketitle

\begin{figure*}[!t]
    \centering
    \includegraphics[width=2\columnwidth]{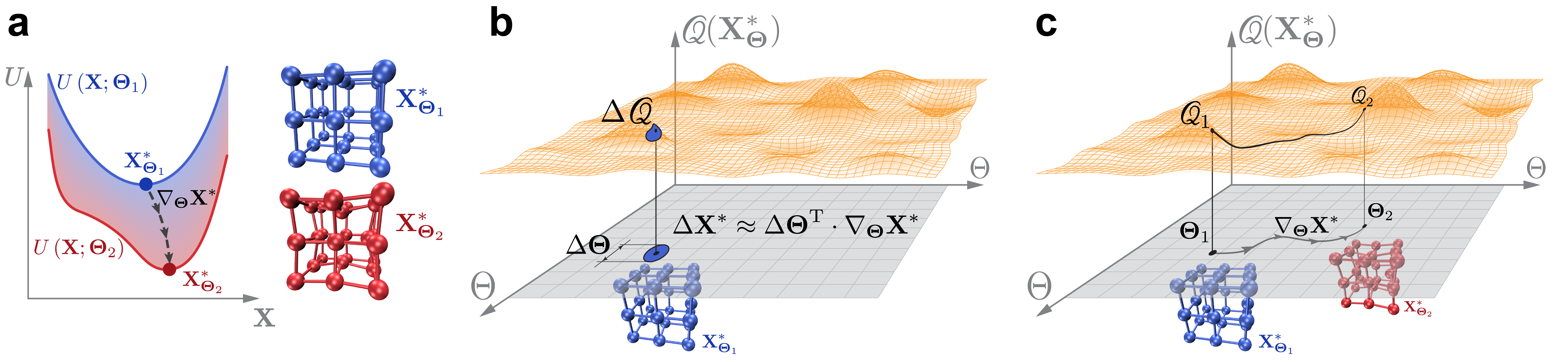}
    \caption{\textbf{Schematic of the implicit derivative approach in atomic systems.} \textbf{a} A local minimum 
    $\X^*_\T$ will change under changes in potential parameters $\T$. 
    The implicit derivative is the gradient of this change, allowing 
    a first-order prediction of structural variations and a second-order prediction of energy variations.
    \textbf{b} Uncertainty quantification application. Having the uncertainty in the parameter space, $\Delta \T$, implicit derivative allows one to compute the uncertainty of any molecular statics property $\mathit{Q}$.
    \textbf{c} Inverse design application. With initial configuration $\X^*_{\T_1}$ and potential $\T_1$, one defines a target configuration $\X^*_{\T_2}$. Implicit derivative allows one to effectively minimize the potential parameters that provide the target configuration.
    }
    \label{fig:Implicit-Der}
\end{figure*}

%%%%%%%%%%%%%%%%%%%%%%%%%%%%%%%%%%%%%%%%%%%%%%%%%%%%%%%%%%%%%%%%%%%%%%%%%%%%
\section{Introduction}
%%%%%%%%%%%%%%%%%%%%%%%%%%%%%%%%%%%%%%%%%%%%%%%%%%%%%%%%%%%%%%%%%%%%%%%%%%%%

Atomic simulations employing interatomic potentials are an essential
tool of computational materials science~\cite{van2020roadmap}. 
Classical models such as Lennard-Jones potentials
are central to the study of glasses and polymer systems, 
whilst modern, data-driven models are becoming quantitative 
surrogates for \textit{ab initio} calculations~\cite{deringer2019machine}.
For solid-state materials, the energy landscape of relaxed atomic geometries is central to exploring thermodynamic, diffusive and mechanical properties
~\cite{swinburne2020automated,grigorev2023calculation,wales_energy_2003,proville2012quantum}.

Regardless of the interatomic potential employed, changing parameters will change any
quantity of interest extracted from a simulation. For simple classical 
models, parameter variation is essential to explore the model's phenomenology~\cite{goodrich2021designing,wales_energy_2003}. 
For modern data-driven models that target quantitative accuracy, 
parameter uncertainties can be estimated by Bayesian regression on 
training data~\cite{goryaeva2021,swinburne2024parameter} and should be 
forward-propagated to simulation results to bound quantities of interest~\cite{musil2019fast,George_2019,Zhou_2019,Ha_2021}. \\

%\red{Any forward propagation scheme must 
%account for the strong correlation between individual energy 
%or force evaluations when calculating e.g. formation energies 
%or dynamical averages, as these will strongly affect (typically \textit{reduce}) uncertainty 
%on the final simulation result.} UQ~\cite{Kulichenko_2023, Wen_2020}, growing interest in high-entropy alloy systems~\cite{George_2019,Zhou_2019,Ha_2021}. inverse design~\cite{Giuntoli_2021,Banik_2023}
%Efficient uncertainty quantification (UQ) has an ever-growing interest for the large-scale simulations. The mac

Backpropagation of structural modifications to changes in parameters is finding application in inverse material design~\cite{Giuntoli_2021,Banik_2023}, training interatomic potentials from experimental data~\cite{thaler2022deep}, or `fine-tuning'
simple interatomic potentials to reproduce desired self-assembly kinetics~\cite{goodrich2021designing}. More recently,
`universal' machine learning potentials have shown near-quantitative accuracy across large portions of the periodic table~\cite{batatia2022mace,batatia2024foundation}. 
This has raised interest in backpropagation for fine-tuning 
universal models for specific applications~\cite{deng2024overcoming}, and opened the possibility of 
targeted design through navigation of a smooth latent composition representation~\cite{nam2024interpolation}.\\

Forward and backpropagation of parameter variation through complex simulations 
is typically achieved using reverse-mode automatic differentiation (AD) routines,
which offer high arithmetic efficiency at the price of large memory requirements
~\cite{schoenholz2020jax,ablin2020super,blondel2022efficient,Liu_2023}.
Whilst AD offers clear advantages in implementation,
their significant memory burden complicates application to large atomic systems,
especially when targeting higher derivatives beyond forces. 
Existing methods to propagate variation 
in parameters to variation in simulation results thus
typically employ resampling: new parameters are drawn from some
distribution~\cite{musil2019fast} and simulations are repeated. 
Whilst conceptually straightforward, backpropagation is not possible, 
assessing convergence is challenging, and the cost can be potentially very large,
especially for simulation results requiring geometry minimization for each potential sample.
\\

In this paper, we explore an alternative approach, {analytically} expanding the structure of relaxed minima to first order in parameter variation, giving a second order expansion of the total energy. The expansion is achieved through evaluation of an 
\textit{implicit} derivative~\cite{krantz2002implicit}, i.e. the 
derivative of a function defined as fixed point, in this case, the atomic structure of a local minimum 
(Fig.~\ref{fig:Implicit-Der}a). 
%The extension to saddle points will be presented in future work.\ivan{-- We say this twice, here and in Results. Could we keep this sentence only in Results?} 
Our main results are that the implicit derivative enables 1) 
forward propagation of parameter uncertainties to simulation results for 
orders of magnitude less computational effort than resampling schemes, 
allowing rapid propagation of parameter uncertainties and 2) backpropagation of 
structural variations to target composition-induced structural rearrangements
in multi-thousand-atom systems, a challenging task for any other approach 
({see illustration of these two ideas in Fig.~\ref{fig:Implicit-Der}b and \ref{fig:Implicit-Der}c}). 

We implement and compare methods to evaluate the implicit 
derivative using AD~\cite{ablin2020super,blondel2022efficient}, 
dense linear algebra and a sparse operator approach, 
using the \texttt{jax-md}~\cite{schoenholz2020jax} and 
\texttt{LAMMPS}~\cite{LAMMPS} simulation codes. 
We find AD routines for the implicit derivative reach GPU memory limits 
for $\sim$1000-atom systems even on best-in-class hardware. 
In contrast, the sparse operator technique reduces to a constrained 
minimization in \texttt{LAMMPS}, allowing memory-efficient and highly 
parallelized evaluation. Our developed approach enables uncertainty quantification 
and inverse design studies with the large atomic simulations essential to 
capture realistic defect structures.

For the purposes of backpropagation, our expansion can be used for any form of interatomic potential. However, for the forward propagation in
uncertainty quantification, the expansion captures parameter variation in the vicinity of some minimum of the loss. In this paper we therefore focus on classical~\cite{schoenholz2020jax,xie2023ultra,xie2023ultra,del2024insights}
and linear-in-descriptor interatomic potentials~\cite{Thompson_snap_2015,allen2021atomic,goryaeva2021,podryabinkin2017active,lysogorskiy2021performant} whose loss typically has a well defined global minimum
rather than the multi-modal loss landscape of neural network potentials~\cite{wales_energy_2003,batatia2022mace,batzner20223}.

The paper is structured as follows. We first define the implicit derivative, the
Taylor expansion approximations used and their evaluation using automatic 
differentiation or linear algebra techniques. We then describe how the implicit 
derivative can be used in forward and backpropagation of parameter variations. 
Forward propagation of parameter variation is demonstrated using classical and machine 
learning potentials to explore lattice distortion and vacancy defect formation~\cite{reali2021macroscopic}.
Finally, we apply the implicit derivative approach to the inverse material design problems, finding parameter variations 
that stabilize subtle solute-induced dislocation core reconstructions in tungsten~\cite{grigorev2023calculation}.

%%%%%%%%%%%%%%%%%%%%%%%%%%%%%%%%%%%%%%%%%%%%%%%%%%%%%%%%%%%%%%%%%%%%%%%%%%%%
\section{Results}
%%%%%%%%%%%%%%%%%%%%%%%%%%%%%%%%%%%%%%%%%%%%%%%%%%%%%%%%%%%%%%%%%%%%%%%%%%%%
\subsection{Implicit derivative of atomic configurations}
\label{sec:impl_der}
%%%%%%%%%%%%%%%%%%%%%%%%%%%%%%%%%%%%%%%%%%%%%%%%%%%%%%%%%%%%%%%%%%%%%%%%%%%%

We consider a system of $N$ atoms in a periodic supercell of volume $V$,
with atomic coordinates $\X\in\mathbb{R}^{N\times3}$ and a supercell 
matrix $\C\in\mathbb{R}^{3\times 3}$, $V={\rm det}(\C)$. 
Changes to the supercell $\C\to\C+\delta\C$ are defined to induce homogeneous deformations, 
as in e.g. energy volume curves. 
We then define scaled atomic coordinates $\tX$, 
such that $\X\equiv\tX\C$ and the tuple $(\tX,\C)$ fully describes atomic configurations 
with periodic boundary conditions.
%and fixed atomic species. 
With a vector of $N_D$ potential parameters $\T$, 
a potential energy model $\oU(\tX, \C; \T)$, has stationary configurations 
$(\tX^*_\T,\C^*_\T)$ satisfying
\begin{equation}
    {\bf\nabla}_{\tX} \oU(\tX^*_\T, \C^*_\T; \T)\equiv{\bf 0}, \quad 
    {\bf\nabla}_\C \oU(\tX^*_\T, {\bf C}^*_\T; \T)\equiv 0
    \label{minima0}
\end{equation}
where $(\tX^*_\T,\C^*_\T)$ is one of the exponentially many stationary points in the energy
landscape~\cite{wales_energy_2003}. In the following, we only consider minima; 
extension to the treatment of saddle points will be presented in future work.
Under a parameter variation $\T + \delta \T$ the scaled positions and supercell matrix 
are defined to change as
\begin{align}
    \tX^*_{\T+\delta\T} &= \tX^*_{\T}  +  \delta\T \nabla_\T \tX_\T^* + \mathcal{O}(\delta\T^2), \\
    \C^*_{\T+\delta\T} &= \C^*_\T + \delta \T \nabla_\T \C^*_\T + \mathcal{O}(\delta\T^2),
\end{align}
where $\nabla_\T \tX_\T^*\in\mathbb{R}^{N_D\times N \times 3}$ and
$\nabla_\T \C^*_\T\in \mathbb{R}^{N_D\times3\times3}$ are \textit{implicit} derivatives that determine how a \textit{stationary} configuration changes with the variation of potential parameters. 
In practical applications, supercell variation $\C$ is typically constrained; 
for simplicity, we will only consider fixed-volume simulations or 
isotropic variations controlled by a homogeneous strain $\epsilon^*_\T\in\mathbb{R}$ around a reference
supercell ${\bf C}_0$
\begin{equation}
\C^*_\T=
\left[1+ \epsilon^*_\T\right]\C_0
,\quad 
\nabla_\T\C^*_{\T}=(\nabla_\T \epsilon^*_\T)\C_0,
\label{eq:isotropic}
\end{equation}
where $\nabla_\T \epsilon^*_\T\in \mathbb{R}^{N_D}$.
Taylor-expanding equations (\ref{minima0}) to first order in $\delta\T$,
it is simple to show that $\nabla_\T \tX_\T^*, \nabla_\T \epsilon_\T^*$
solve the system of linear equations
\begin{align}
    \left[
    %\begin{matrix}
    %\nabla_\T \tX_\T^* \\
    %\nabla_\T \epsilon_\T^*
    %\end{matrix}
    % AN ALTERNATIVE
    \begin{matrix}
    \nabla_\T \tX_\T^* & \nabla_\T \epsilon_\T^*
    \end{matrix}
    \right]
    \left[
    \begin{matrix}
    \nabla^2_{\tX\tX}\oU & \nabla^2_{\tX\epsilon}\oU\\
    \nabla^2_{\tX\epsilon}\oU^\top & \nabla^2_{\epsilon\epsilon}\oU
    \end{matrix}
    \right]
    &=-
    \left[
    \begin{matrix}
    \nabla^2_{\T\tX}\oU \\
    \nabla^2_{\T\epsilon}\oU
    \end{matrix}
    \right],
    \label{eq:impl_der_matrix}
\end{align}
where $\nabla^2_{\tX\tX}\oU$ is the Hessian matrix in scaled coordinates, $\nabla^2_{\epsilon\epsilon}\oU$ is proportional to the bulk modulus of the system, $-\nabla^2_{\epsilon\tX}\oU$ is proportional to change in atomic forces under a homogeneous strain and $\nabla^2_{\T\tX}\oU$, $\nabla^2_{\T\epsilon}\oU$ are mixed curvatures. Whilst the solution of~(\ref{eq:impl_der_matrix}) in principle requires $\mathcal{O}(N^3)$ effort due to the Hessian, we introduce a Hessian-free solution method below, allowing application to large systems.\\

%%%%%%%%%%%%%%%%%%%%%%%%%%%%%%%%%%%%%%%%%%%%%%%%%%%%%%%%%%%%%%%%%%%%%%%%%%%%
\subsection{Taylor expansion of stationary energies and volumes using implicit derivatives}
%%%%%%%%%%%%%%%%%%%%%%%%%%%%%%%%%%%%%%%%%%%%%%%%%%%%%%%%%%%%%%%%%%%%%%%%%%%%

In our numerical experiments, we will compare three levels of approximate solution to the linear equations (\ref{eq:impl_der_matrix}): 
\begin{itemize}
    \item \textit{constant (c)}: $\nabla_\T \epsilon_\T^*=0$, $\nabla_\T \tX_\T^*={\bf 0}$
    \item \textit{homogeneous (h)}: $\nabla_\T \epsilon_\T^*\neq0$ , $\nabla_\T \tX_\T^*={\bf 0}$
    \item \textit{inhomogeneous (ih)}: $\nabla_\T \epsilon_\T^*=0$ , $\nabla_\T \tX_\T^*\neq{\bf 0}$,
\end{itemize}
with the full expansion then given the shorthand \textit{h+ih}. 
Under changes in parameters $\T$, changes in the stationary energy and volume (or equivalently strain) admit the implicit Taylor expansions
\begin{align}
    \delta^{(\zeta)}\oU^* 
    &\equiv\delta \T \nabla_{\T} \oU
    + \delta \T {\bf H}_{\zeta} \delta \T^\top + \mathcal{O}(\delta\T^3)
    \label{eq:energy}
    \\
\delta^{(\zeta)}\epsilon^*
&\equiv
\delta \T \nabla_{\T}\epsilon^{*}_{\zeta}+\mathcal{O}(\delta\T^2)
,
\label{eq:strain}
\end{align}
where 
$ {\bf H}_{\zeta}\in\mathbb{R}^{N_D\times N_D}$ is a generalized curvature within a given level of approximation and $\zeta={c,h,ih,h+ih}$ is the level of approximation used. Expressions for $\nabla_{\T}\epsilon^{*}_{\zeta}$ and ${\bf H}_{\zeta}$ are given in the supplementary material.

In all three approaches, the potential parameters $\T$ vary; therefore, each method predicts changes in energy.
However, only $\zeta={h,ih,h+ih}$ predict changes in structure. For constant volume relaxations, the inhomogeneous $\zeta=ih$ expansion is asymptotically exact. For variable volume relaxations, the full $\zeta=h+ih$ expansion is asymptotically exact, but as we show below, the cheaper homogeneous $\zeta=h$ expansion can also give accurate results when the focus is on changes to the energy or volume, rather than structure. 

%%%%%%%%%%%%%%%%%%%%%%%%%%%%%%%%%%%%%%%%%%%%%%%%%%%%%%%%%%%%%%%%%%%%%%%%%%%%
\subsection{Evaluation of the implicit derivative through sparse and dense linear algebra methods}
%%%%%%%%%%%%%%%%%%%%%%%%%%%%%%%%%%%%%%%%%%%%%%%%%%%%%%%%%%%%%%%%%%%%%%%%%%%%

In the linear equations (\ref{eq:impl_der_matrix}), the $\nabla^2_{\epsilon\epsilon}U$ and $\nabla^2_{\epsilon\tX}U$ derivatives require only a few $\mathcal{O}(N)$ force calls for evaluation. As a result, the \textit{homogeneous} approximation requires minimal computational effort, but all knowledge of structural changes is missing as $\nabla_\T\tX^*$ is not evaluated.
Evaluation of $\nabla_\T\tX^*$ for the \textit{inhomogeneous} approach requires $\mathcal{O}(N^2)$ finite difference evaluations of the Hessian matrix $\nabla^2_{\tX\tX}U$ and $\mathcal{O}(N^3)$ solution of the dense linear equation (\ref{eq:impl_der_matrix}). 
Whilst of reasonable cost for small systems ($N<2000$), 
study of extended defects where $10^4<N<10^6$ requires significant, typically prohibitive, computational resources and careful use of shared memory parallel linear algebra techniques~\cite{proville2012quantum}. \\

To overcome this limitation, we note that the Hessian matrix is highly sparse due to the strong locality of atomic forces. In this regime, efficient solutions of the linear equations (\ref{eq:impl_der_matrix}) can be obtained using iterative algorithms. In addition, such algorithms do not require access to every element of the Hessian at each iteration, only a linear operator that gives the action of the Hessian on some vector ${\bf V}\in\mathbb{R}^{N\times3}$, i.e. 
$\mathcal{L}({\bf V})={\bf V}\nabla^2_{\tX\tX}U$.

Avoiding direct Hessian evaluation can give a much faster time-to-solution. We define the operator
\begin{equation}
    \mathcal{L}({\bf V})
    \equiv
    \lim_{\alpha\to0}
    {\bf\nabla}_{\tX} \oU(\tX^*_\T+\alpha{\bf V}, \C^*_\T; \T)/\alpha
    ,
    \label{lin_op}
\end{equation}
which, in the limit, is equal to $\mathcal{L}({\bf V})={\bf V}\nabla^2_{\tX\tX}\oU$ as desired and only requires $\mathcal{O}(N)$ force calls for evaluation. To compute the \textit{inhomogeneous} implicit derivative, we apply the sparse linear operator to each vector $[\nabla_\T \tX_\T^*]_l\in\mathbb{R}^{N\times3}$, $l\in[1,N_D]$. Details of 
our $\mathcal{O}(NN_D)$ massively parallel method to compute ${\bf\nabla}_\T\tX^*$ in \texttt{LAMMPS}~\cite{LAMMPS} at finite values of $\alpha$ is described in methods section \ref{sec:methods-sparse}. In section \ref{sec:inverse-design}, we apply the method to enable the use of the implicit derivative in large atomic systems.

%%%%%%%%%%%%%%%%%%%%%%%%%%%%%%%%%%%%%%%%%%%%%%%%%%%%%%%%%%%%%%%%%%%%%%%%%%%%
\subsection{Evaluation of the implicit derivative with automatic differentiation methods}
%%%%%%%%%%%%%%%%%%%%%%%%%%%%%%%%%%%%%%%%%%%%%%%%%%%%%%%%%%%%%%%%%%%%%%%%%%%%
AD-enabled simulation schemes such as \texttt{jax-md}~\cite{schoenholz2020jax} 
can clearly evaluate all terms in equation (\ref{eq:impl_der_matrix}) or the sparse operator 
approach (\ref{lin_op}). Recently, implicit differentiation schemes have been implemented in 
\texttt{jaxopt}~\cite{ablin2020super,blondel2022efficient}, allowing 
direct evaluation of e.g., ${\bf\nabla}_\T\tX^*$ by differentiating through 
the minimization algorithm chosen in \texttt{jax-md}. We have implemented and tested 
all approaches in AD for the binary Lennard-Jones system described below.
Despite the simplicity of the potential form, using AD implicit derivative schemes incurs extremely large memory usage, reaching the 80GB limit on NVIDIA A100 
GPUs for only a few thousand atoms, as we detail in the supplementary material. 
We thus conclude that existing AD schemes for direct evaluation of the implicit derivative or Hessian matrices 
are ill-suited for application to the thousand-atom systems essential for many materials science problems.
In contrast, our sparse operator technique has the same memory usage as any structural minimization. 
Whilst still incurring a significantly greater memory burden than non-AD methods implemented in \texttt{LAMMPS}, 
our sparse operator is ideal for implicit derivative evaluation in AD-enabled 
schemes. 
%Investigation of how (\ref{lin_op}) can be used with neural network-based interatomic 
%potentials\cite{batatia2022mace} is left for a future study.

\begin{figure}[t!]
    \centering
    \includegraphics[width=1.0\columnwidth]{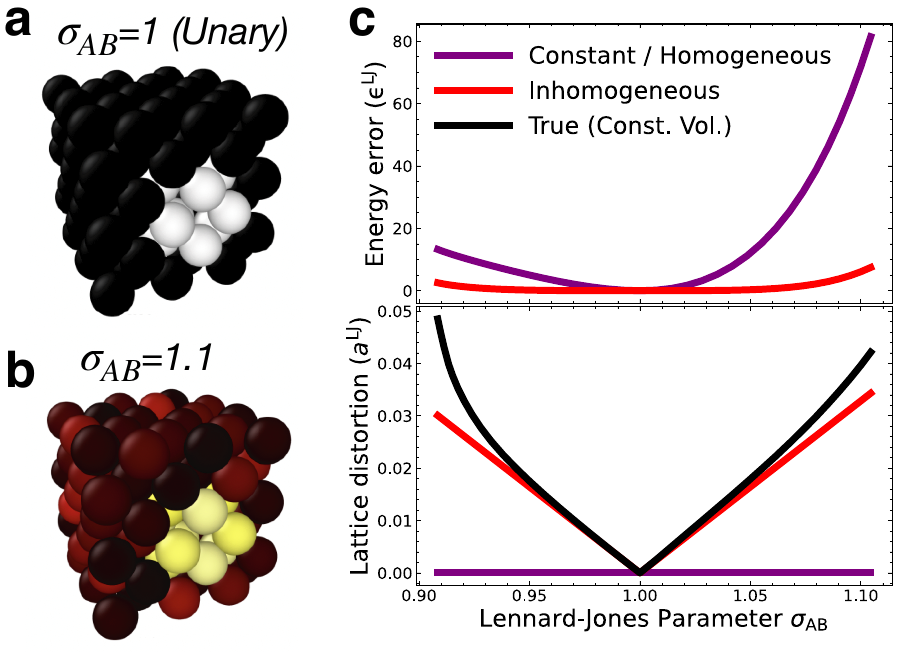}
    \caption{
    \textbf{Automatic implicit differentiation with classical potentials.}
    \textbf{a}, \textbf{b} Atomic configurations of a single vacancy configuration with $\sigma_{\rm AB}=1$ (unary) and $\sigma_{\rm AB}=1.1$ (random binary). Atom colors encode fcc centrosymmetry in the range $[0,1.25]$; black color corresponds to perfect fcc structure. \textbf{c} Relaxed energy and lattice distortion for $\sigma_{\rm AB}\in[0.9, 1.1]$ at fixed volume.
    %where \textit{constant} and \textit{homogeneous} approximations are identical.
    %The \textit{inhomogeneous} expansion accounts for changes in $\tX$, giving much more accurate energy predictions. 
    The true values are obtained with the energy minimization at each $\sigma_{\rm AB}$ value.
}
    \label{fig:LJ}
\end{figure}

\begin{figure*}[t!]
    \centering
    \includegraphics[width=2.0\columnwidth]{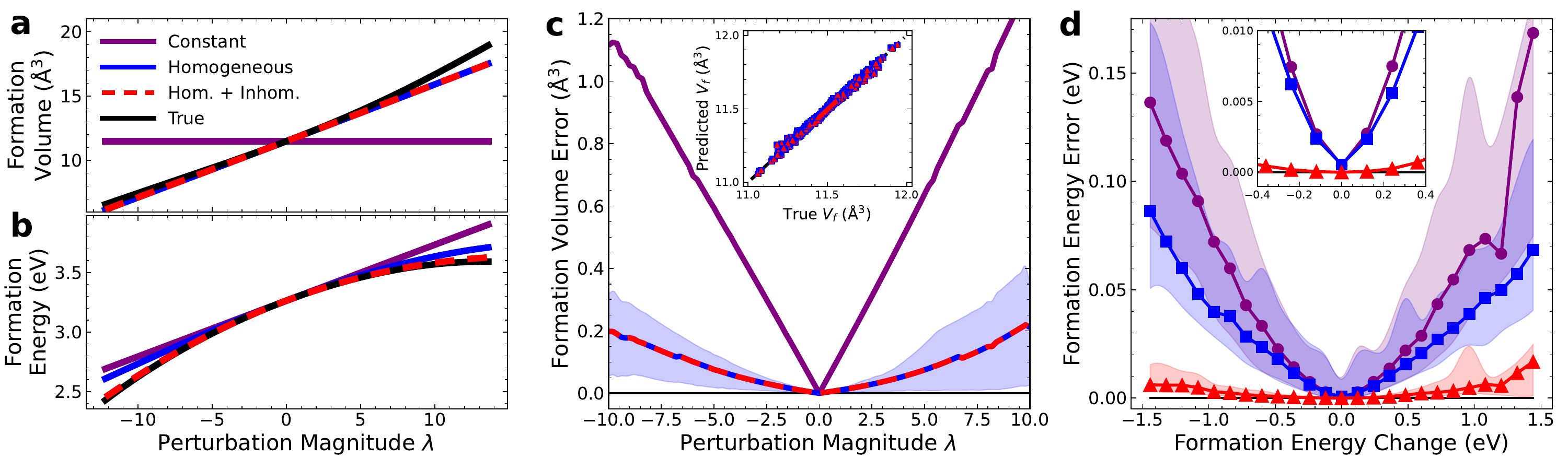} 
    \caption{\textbf{Implicit differentiation of vacancy defect formation with \texttt{SNAP} potentials.}
    \textbf{a}, \textbf{b} Formation volume and formation energy vs potential perturbation magnitude for one of 100 tungsten potential samples used in the study.
    \textbf{c} Average absolute error of formation volume vs perturbation magnitude. On the inset, predicted vacancy formation volume vs its true value obtained with the full relaxation.
    \textbf{d} Error vs change in vacancy formation energy averaged over the set of potentials and perturbation magnitudes. On the inset, zoom into a smaller formation energy range.
    In \textbf{c},\textbf{d} the shaded regions indicates the [16\%, 84\%] percentile ($\pm1\sigma$) range across all parameter variations.
    Line colors across the panels indicate the prediction method with the legend presented in \textbf{a}. In \textbf{d}, symbols are added at data points. Pristine supercell size of 128 atoms was used in the study.
    }
    \label{fig:potential-samples}
\end{figure*}

%%%%%%%%%%%%%%%%%%%%%%%%%%%%%%%%%%%%%%%%%%%%%%%%%%%%%%%%%%%%%%%%%%%%%%%%%%%%
\subsection{Backpropagation of parameter variations for inverse design problems}
\label{sec:implicit-loss}
%%%%%%%%%%%%%%%%%%%%%%%%%%%%%%%%%%%%%%%%%%%%%%%%%%%%%%%%%%%%%%%%%%%%%%%%%%%%

Inverse design aims to produce materials with specified (desirable) properties
through inverting structure-property relationships. 
However, this typically requires high-throughput searches, which necessarily cannot afford to perform atomistic simulations of e.g. defect structures and mechanisms to predict mechanical properties. 
The implicit derivative can be used to make a first step toward gradient-led inverse design, 
allowing us to find interatomic parameters $\T$ that stabilise some atomic configuration 
$\X^{*}_{\rm ref}$ observed in an \textit{ab initio}-accurate simulation e.g. using the density-functional theory (DFT)~\cite{grigorev2023calculation}. 
%With $\X^*_{\T}$ being the local minimum found when 
%minimizing $U(\X;\T)$ starting from $\X^{*}_{\rm ref}$, 
We define an implicit loss function:
\begin{equation}
    %L(\T) = \frac{1}{2}\, \big[\X^*_{\T} - \X^{*}\big]:\big[\X^*_{\T} - \X^{*}\big]
    L(\T) = \frac{1}{2} \| \X^*_{\T} - \X^{*}_{\rm ref} \|^2.
    \label{eq:implicit-loss}
    % we only use this for one config at a time
\end{equation}
%where $\X^*_t$ is the target system configuration.
%where we use the notation $:$ to indicates summation over the $N$ atomic sites and $3$ spatial indices.
To effectively minimize the loss, the loss gradient is required, and the implicit derivative is necessary to compute the derivative of the loss via the chain rule:
\begin{equation}
    \label{eq:implicit-loss_derivative}
    \nabla_{\T} L(\T) = 
    \nabla_{\T}\X^*_{\T} \cdot %: 
    \big[
    \X^*_{\T} - \X^{*}_{\rm ref}
    \big]\in\mathbb{R}^{N_D}.
\end{equation}
%\textbf{Future work will investigate incorporating implicit loss functions (equation \ref{eq:implicit-loss})
%into an inverse design framework by associating $\T$ with composition, a challenging
%task which is beyond the scope of the present study.} \ivan{-- Move to Conclusion?}
An immediate application of the implicit loss, equation (\ref{eq:implicit-loss}), is the ability to `fine-tune', or retrain~\cite{grigorev2023calculation}, interatomic potentials to reproduce important DFT minima. Fine-tuning has gained increasing interest following the rise of `universal' machine learning potentials~\cite{batatia2024foundation}. %which are beyond the scope of this study. 
In practice, one typically includes the loss against the original training database to
regularize the fine-tuning fit. However, we have found that minimizing (\ref{eq:implicit-loss}) 
in practice produces only very small perturbations to the final potential.

The ability to fine-tune interactions is of particular relevance for low energy structures such as dislocation lines~\cite{goryaeva2021,grigorev2023calculation}, which typically require careful weighting in potential fitting~\cite{batatia2024foundation}. In section~\ref{sec:inverse-design}, we use implicit loss minimization to find 
solute substitutions which induce `hard' screw dislocation 
core reconstruction in tungsten, and fine-tune a tungsten-beryllium potential 
to correctly reproduce \textit{ab initio} observations.

%%%%%%%%%%%%%%%%%%%%%%%%%%%%%%%%%%%%%%%%%%%%%%%%%%%%%%%%%%%%%%%%%%%%%%%%%%%%
\subsection{Prediction of the total relaxed energy in classical potentials with automatic differentiation}
%%%%%%%%%%%%%%%%%%%%%%%%%%%%%%%%%%%%%%%%%%%%%%%%%%%%%%%%%%%%%%%%%%%%%%%%%%%%

The binary Lennard-Jones potential is a classical model for nanoclusters and glassy systems~\cite{wales_energy_2003}. 
The model is defined by six parameters,
$\T=[\epsilon^{\rm LJ}_{\rm AA},\epsilon^{\rm LJ}_{\rm AB},\epsilon^{\rm LJ}_{\rm BB},\sigma_{\rm AA},\sigma_{\rm AB},\sigma_{\rm BB}]$, with a total energy
\begin{equation}
    \oU(\tX,\C;\T)
    =
    \sum_i
    \sum_{j\in N_i} 
    \epsilon^{\rm LJ}_{s_is_j}
    \left(
    \frac{\sigma^{12}_{s_is_j}}{r^{12}_{ij}}
    -
    \frac{\sigma^{6}_{s_is_j}}{r^{6}_{ij}}
    \right),
\end{equation}
where $s_i$ is species $i$, $s_i\in[A,B]$, $r_{ij}$ is the minimum image distance (as determined by $\C$) between atoms $i,j$ and $N_i$ is the set of neighbors of $i$. As discussed above, {automatic differentiation enabled by} \texttt{jax-md} was used to 
study this simple system, with all examples shown using the dense linear algebra approach 
to evaluate the implicit derivative at constant volume.

To simplify the problem, we set $\epsilon^{\rm LJ}_{\rm AA}=\epsilon^{\rm LJ}_{\rm AB}=\epsilon^{\rm LJ}_{\rm BB}$ and $\sigma_{\rm AA}=\sigma_{\rm BB}=1$, leaving $\Theta=\sigma_{\rm AB}$ as the only varying parameter in this example. 
When $\sigma_{\mathrm{AB}}=1$, all atoms are identical and the system is a unary fcc lattice; we 
additionally remove one atom to form a vacancy and promote additional deformation. When 
$\sigma_{\mathrm{AB}}\neq1$ the system becomes a random fcc binary alloy, with lattice distortion in 
the bulk and around the vacancy (see Fig.~\ref{fig:LJ}a and \ref{fig:LJ}b). 
Figure \ref{fig:LJ}c shows the \textit{inhomogeneous} implicit derivative around $\sigma_{AB}=1$ that gives an excellent prediction of the total energy and lattice distortion for 
$\sigma_{AB}\in[0.95,1.05]$, with mild disagreement as $|\sigma_{AB}-1|$ grows. At constant
volume, both \textit{constant} and \textit{homogeneous} approximations are equivalent
and predict no structural change, with significantly higher errors. 
In the remainder of this paper, we focus on modern machine learning potentials.
%
%
%%%%%%%%%

%%%%%%%%%%%%%%%%%%%%%%%%%%%%%%%%%%%%%%%%%%%%%%%%%%%%%%%%%%%%%%%%%%%%%%%%%%%%
\subsection{Prediction of defect formation energies and volumes with machine learning potentials}
\label{sec:vac_form}
%%%%%%%%%%%%%%%%%%%%%%%%%%%%%%%%%%%%%%%%%%%%%%%%%%%%%%%%%%%%%%%%%%%%%%%%%%%%

Almost all modern interatomic models use high-dimensional regression techniques from the machine learning community~\cite{bartok2010,Bartok_machine_2017,goryaeva2021,batatia2024foundation}. A common first step is to represent atomic environments as $N_D$ per-atom descriptor functions $D_l(\tX,\C,i)$, $l\in[1,N_D]$ which describe the atomic environment around an atom of index $i$. In practice, the descriptor functions also have species-dependent hyperparameters which must also be tuned, but in the following, we assume that these
are fixed. We use the widely implemented \texttt{SNAP} Bispectrum descriptors~\cite{Thompson_snap_2015} 
with $N_D=55$ (see methods, section \ref{sec:methods-QM-ML}, for multi-specie generalization), giving a potential energy
\begin{equation}
    U(\tX,\C;\T) = \T\sum_i{\bf D}(\tX,\C,i) \equiv \T\cdot\D(\tX,\C),
    \label{eqn:snap}
\end{equation}
where $\T\in\mathbb{R}^{N_D}$ is the potential parameter vector
and $\D(\tX,\C)\in\mathbb{R}^{N_D}$ is the total descriptor vector. 
The advantage of the linear functional form is that $\nabla_{\T} U={\bf D}$ and $\nabla^2_{\T\tX}U=\nabla_{\tX}{\bf D}$, required for the solution of equation~(\ref{eq:impl_der_matrix}), are readily evaluated without any numerical or automatic differentiation schemes. 
%\textbf{Future work will apply the implicit expansion to more complex interatomic potentials, e.g. graph neural networks}~\cite{batatia2022mace} \ivan{-- Move to Conclusion?}.

For multi-scale uncertainty quantification study, we use DFT training data for tungsten from Goryaeva \textit{et al.}~\cite{goryaeva2021} and 
generate 100 samples $\T_m$ ($m$ is a sample index) from a parameter distribution using the approach described in \cite{swinburne2024parameter}. 
%The potential samples are suitable for multi-scale uncertainty quantification. 
We are primarily interested in using the variations of parameter samples 
away from the reference potential $\bar{\T}$ to test our implicit expansion method. 
To this end, we define an additional `perturbation magnitude' $\lambda$ and generate samples with
\begin{equation}
    %\T = (1-\lambda)\bar{\T} + \lambda\T_l,
    \T(\lambda,m) = \bar{\T} + \lambda(\T_m - \bar{\T}),
\end{equation}
such that $\lambda=0$ corresponds to the reference potential 
$\bar{\T}$ and $\lambda=1$ corresponds to the original sample $\T_m$. 
We then generate very large (and thus sometimes unphysical) 
perturbations with $\lambda\in[-25,25]$, with a step $\Delta \lambda=0.2$, truncating only when the bcc 
lattice became unstable. This yielded a total ensemble of around $20000$ stable potentials.
For each potential, we calculated the formation vacancy defect~\cite{AshcroftMermin1976}, allowing for relaxation of
both structure and volume, meaning that only the full \textit{ih+h} expansion
is expected to be asymptotically exact. The diversity of the resultant dataset allows for 
a robust test of our implicit Taylor expansions (\ref{eq:energy}) and (\ref{eq:strain}). 

\begin{figure*}[t!]
    \centering
    \includegraphics[width=2\columnwidth]{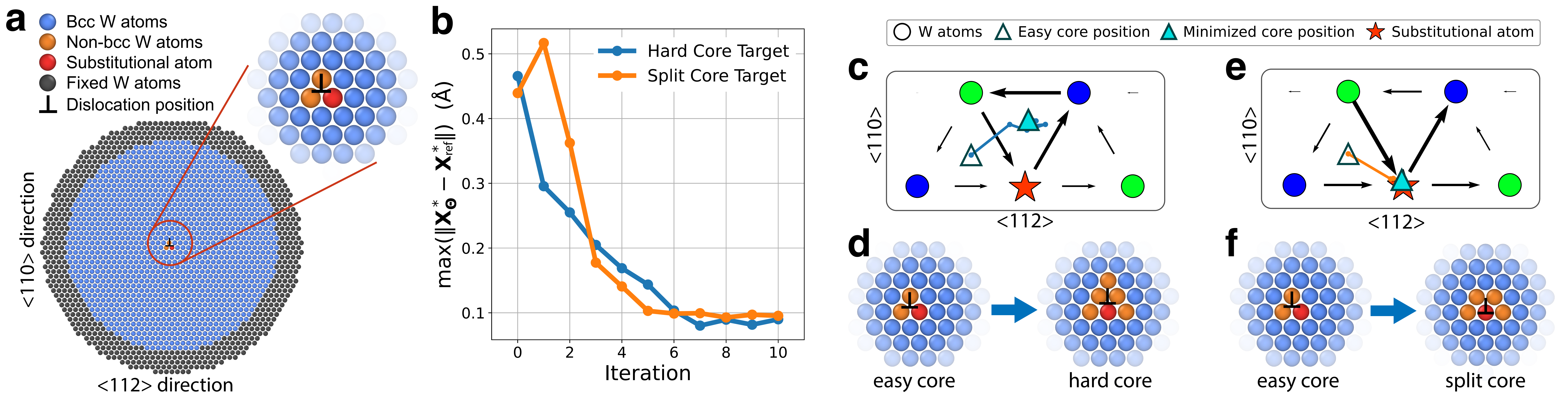}
    \caption{\textbf{Targeting dislocation core reconstruction in tungsten.} \textbf{a} Dislocation cluster supercell. The $\sim$2000-atom cell has fixed boundary conditions in the $\langle110\rangle$ and $\langle112\rangle$ directions and periodic boundary conditions in the $\langle111\rangle$ direction with one Burgers vector within the cell. The `substitutional' atom at the dislocation core (in red), initially has interaction parameters corresponding to tungsten, which vary during the loss minimization procedure. Around 500 atoms are fixed at the cluster border to stabilize the dislocation core structure~\cite{grigorev2023calculation}. Non-bcc atoms are identified using the \texttt{OVITO} package~\cite{Stukowski_2009}.
    \textbf{b} Maximal difference of atomic positions between a current and the target configurations during the loss minimization procedure for target configurations set to `hard' and `split' cores.
    \textbf{c}, \textbf{e} Differential displacement maps~\cite{vitek1974theory} corresponding to final configurations of the minimization. The starting configuration for both simulations is the `easy' core and the target configurations are the `hard' core (\textbf{c}) and `split' core (\textbf{e}). Transient dislocation core positions during the minimization are presented with lines starting at empty and ending at solid triangles, that depict initial and final core positions.
    \textbf{d}, \textbf{f} Atomic configurations around the dislocation core with `hard' core (\textbf{d}) and `split' core (\textbf{f}) target configurations.
    }
    \label{fig:WX}
\end{figure*}

Figures~\ref{fig:potential-samples}a and \ref{fig:potential-samples}b illustrate this approach. 
As expected, this form of perturbation applied to all stable potentials produced a wide range of very strong perturbations. Fig.~\ref{fig:potential-samples}c shows the formation volume error across the samples as a function of perturbation magnitude $\lambda$. Both homogeneous (\textit{h}) and full (\textit{h+ih}) expansions provide a nearly-perfect predictions of vacancy formation volume. However, for the formation energy, the full approach is notably more accurate, with errors under 2\% across a wide (3~eV) range of formation energies (see Fig.~\ref{fig:potential-samples}d). This indicates that whilst the efficient homogeneous 
expansion offers a useful prediction of energy and volume changes, the inhomogeneous term 
allows for most accurate prediction at small to moderate perturbations. 
This asymptotic accuracy is particularly important when using the implicit derivative 
to solve inverse problems, which we discuss in the next section.
%%%%%%%%%%%%%%%%%%%%%%%%%%%%%%%%%%%%%%%%%%%%%%%%%%%%%%%%%%%%%%%%%%%%%%%%%%%%
\subsection{Implicit loss minimization applied to solute-induced dislocation core reconstruction}
\label{sec:inverse-design}
%%%%%%%%%%%%%%%%%%%%%%%%%%%%%%%%%%%%%%%%%%%%%%%%%%%%%%%%%%%%%%%%%%%%%%%%%%%%

%\subsubsection{Dislocation core stability with inverse design}
In this final section we employ the implicit derivative concept to solve a challenging inverse problem: with a starting potential $\T$, and some stationary configurations $\X^*_\T$, we search for the potential parameters that stabilize a structure as close as possible to some target configuration $\X^*_{\rm ref}$. 
%As discussed in section~\ref{sec:implicit-loss} this has application for potential fine-tuning, as we demonstrate below. %More generally, 
The ability to find parameters that yield certain desired structures 
represents a first step towards a range of inverse design strategies, 
in particular, given the ability of emerging universal potential models to 
smoothly interpolate across chemical space~\cite{nam2024interpolation}.
% A deeper exploration of this alchemical exploration will be studied in future work.

To demonstrate minimization of the implicit loss function (equation (\ref{eq:implicit-loss})), we selected a computationally challenging system of a $\sim$2000-atom tungsten disk with a $\langle111\rangle/2$ screw dislocation along the disk axis, as illustrated on Fig.~\ref{fig:WX}a. 
%As detailed in~\citep{grigorev2023calculation}, the outer layers of atoms in the disk are fixed to displacement from elasticity theory and we impose periodic boundary conditions along the dislocation line direction. 
Using initial potential parameters $\T_0$ for W from Goryaeva \textit{et al.}~\cite{goryaeva2021}, the dislocation core relaxes into the `easy' core structure in agreement with \textit{ab initio}.
%, as illustrated in Fig.~\ref{fig:WX}a. 
For the inverse problem, we assign one `alchemical'
atom (the red atom in Fig.~\ref{fig:WX}a) with it's own independent set of parameters $\T$, initially set to $\T_0$.
We use a simple form for the linear multi-specie potential, detailed in the methods, section~\ref{sec:methods-multispec}. 
Modifying the alchemical potential parameters $\T$ whilst keeping $\T_0$ fixed, we aim to stabilize the `hard' and `split' core structures, which are unstable for W~\cite{Grigorev_2020,goryaeva2021}. 
%We generated the target core configurations with the \texttt{matscipy.dislocation}~\cite{Grigorev2024} Python package.
%As the target structures are for pure, single-element W we do not expect the structure induced by the alchemical solute to give an exact match, but we can monitor the effective dislocation core position using the strain-matching approach detailed in~\cite{Grigorev2024}.
We monitor the effective dislocation core position during the minimization using the strain-matching approach detailed in~\cite{grigorev2023calculation}.
The implicit loss minimization is achieved through the procedure presented in Algorithm~\ref{alg:inverse-design}, where the loss gradient is computed according to equation~(\ref{eq:implicit-loss_derivative}). We employ an adaptive step size $h$ as detailed in supplementary material.

\begin{algorithm}[h!]
\caption{Implicit Loss Minimization}\label{alg:inverse-design}
\begin{algorithmic}

\State $\T^{(0)} \gets $ initial potential
\State $\X^*_{\T^{(0)}} \gets $ initial relaxed positions
\State $k \gets 0$
\While{$k\le \mathrm{max\_interations}$}
\State Compute implicit derivative $\nabla_{\T} \X^{*}_{\T^{(k)}}$
\State Compute $\nabla_{\T} L(\T^{(k)})$
\State $\T^{(k+1)} \gets \T^{(k)} - h \nabla_{\T} L(\T^{(k)})$
\State $\X^*_{\T^{(k+1)}} \gets \X^*_{\T^{(k)}} + (\T^{(k+1)} - \T^{(k)}) \nabla_{\T} \X^{*}_{\T^{(k)}}$
\If{ $\|\X^*_{\T^{(k+1)}} - \X^*_{\T^{(k)}}\|_\infty<\varepsilon$}     
    \State \Return $\X^*_{\T^{(k+1)}}$, $\T^{(k+1)}$
\EndIf
\EndWhile
\end{algorithmic}
\end{algorithm}

%i.e. we exit the minimization when the maximum component change in 
%the atomic positions is below some threshold. 

Figure~\ref{fig:WX}b shows the maximal deviation of atomic positions at iteration $k$ for two target structures. The minimization error decreases significantly at first steps and saturates at $\sim$10 iterations for both structures. The error does not reach zero, which we attribute to the target configurations being derived from pure tungsten systems, whereas our minimization involves systems that more closely represent substitutional defects. However, the minimization goals are clearly achieved as seen in Figure~\ref{fig:WX} panels c-f: the `hard' core (Fig.~\ref{fig:WX}c,d) and `split' core (Fig.~\ref{fig:WX}e,f) are located at expected positions denoted by solid triangles.

\begin{figure}[t!]
    \centering
    \includegraphics[width=1\columnwidth]{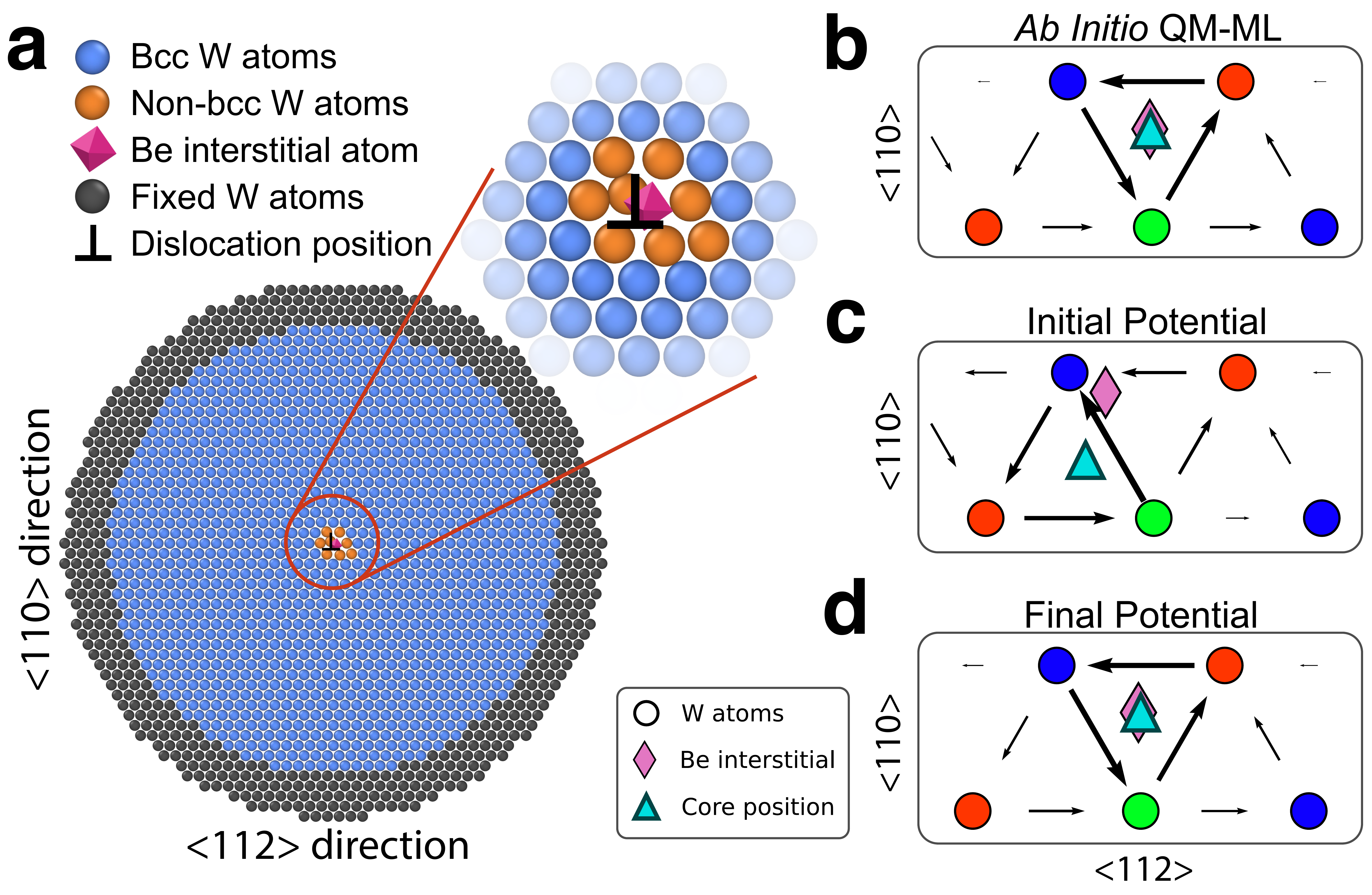}
    \caption{\textbf{`Fine-tuning' Be-induced dislocation core reconstruction in tungsten.} \textbf{a} Dislocation cluster supercell similar to that shown in figure \ref{fig:WX}, but with a Be interstitial in the center of the cluster and no substitution. \textbf{b}, \textbf{c}, \textbf{d} differential displacement maps~\cite{vitek1974theory} with dislocation core and interstitial positions corresponding to the QM-ML configuration (panel \textbf{b}), initial \texttt{SNAP} potential (panel \textbf{c}), and the minimum of the `fine-tuned' potential (panel \textbf{d}).  
    }
    \label{fig:WBe}
\end{figure}

As a final application, we show how the implicit loss minimization can be 
used to fine-tune an initial interatomic potential to match 
\textit{ab initio} training data. Here, our target is a solute-induced 
reconstruction of the $\langle111\rangle/2$ screw dislocation core
caused by an interstitial Be atom, using the same disk geometry as described above and shown in Fig.~\ref{fig:WBe}a. The target data was generated using the
QM/ML simulation method which embeds a DFT region at core as detailed in methods \ref{sec:methods-QM-ML}. As shown in Fig.~\ref{fig:WBe}b, we see that 
Be induces reconstruction to the `hard' core structure, with the Be interstitial sitting at the center of the dislocation. 

Using W-Be \textit{ab initio} training data from Wood \textit{et al.}~\cite{wood2019}, we created an initial set of Be 
interaction parameters $\T$ using the same linear multi-species interatomic potential as above (see methods, section \ref{sec:methods-multispec}) with W parameters
$\T_0$ set to those from Goryaeva \textit{et al.}~\cite{goryaeva2021}. 
The relaxed structure using the initial potential fit is shown in 
Fig.~\ref{fig:WBe}c. It can be seen that in contrast to the QM/ML target, 
the dislocation core remains in the `easy' configuration, with the Be atom lying outside of the central core region. 
Using the procedure described above, we performed implicit loss minimization using this initial relaxed configuration. The implicit loss achieved near-perfect reproduction of the core reconstruction in around 20 iterations (see Fig.~\ref{fig:WBe}d and supplementary material).
%Future work will investigate more sophisticated minimization schemes than the simple gradient descent used here. 

%%%%%%%%%%%%%%%%%%%%%%%%%%%%%%%%%%%%%%%%%%%%%%%%%%%%%%%%%%%%%%%%%%%%%%%%%%%%
\section{Discussion}
%%%%%%%%%%%%%%%%%%%%%%%%%%%%%%%%%%%%%%%%%%%%%%%%%%%%%%%%%%%%%%%%%%%%%%%%%%%%

In this paper, we have investigated the use of the implicit derivative of the relaxed 
atomic structure with interatomic potential parameters, giving a first-order implicit 
Taylor expansion for the relaxed structures and second-order for relaxed energies. 
We detailed how the implicit derivative could be calculated using dense linear algebra, 
requiring Hessian evaluation, automatic differentiation, or a Hessian-free 
linear operator approach, which reduces to a constrained minimization in \texttt{LAMMPS}, allowing
application to arbitrarily large systems.

The implicit expansion enables very efficient forward propagation of parameter uncertainties to simulation results, including the effect of geometry relaxation, essential to capture changes in structure.
%such as strain.
This was demonstrated on simple classical models and machine learning models for pure tungsten. The implicit expansion was able to 
capture a wide range of changes in energy and structure, far beyond 
typical variations associated with potential parameter uncertainty. 
%A forthcoming publication will demonstrate the implicit expansion in a wide-ranging uncertainty quantification study. 
Beyond uncertainty quantification,  our results show that the implicit expansion can also be used to rapidly explore the parameter space of high dimensional interatomic models, 
permitting parametric studies that would be intractable with 
standard methods. In future work, we will explore how the implicit expansion 
can be used in a correlative study of defect structures and implications for
both uncertainty quantification and materials design goals. 

In addition to the forward propagation enabled by the implicit expansion, 
we also investigated the use of the implicit derivative in backpropagation of structural changes to changes in potential
parameters. Our exploratory applications focused on solute-induced dislocation core reconstruction in tungsten, a key feature in understanding plasticity and irradiation damage in bcc metals~\cite{hachet2020screw}. 
We showed how the implicit derivative could be used to `fine-tune'
parameters from an initial fit against training data for tungsten-beryllium, in order to stabilize
the structure seen in \textit{ab initio} calculations. 

In a first effort towards targeted `alchemical' design applications, we used the implicit derivative to find substitutional solute interaction parameters 
which stabilized `hard' or `split' dislocation cores in pure tungsten. The success of 
this effort extends the scope of alchemical machine learning to large-scale
simulations essential for mechanistic studies of e.g. plasticity. We 
anticipate that both approaches will gain ever-increasing application 
with the advent of general-purpose `universal' machine learning potentials,
which will be a focus of future efforts. 

%%%%%%%%%%%%%%%%%%%%%%%%%%%%%%%%%%%%%%%%%%%%%%%%%%%%%%%%%%%%%%%%%%%%%%%%%%%%
\section{Methods}
\label{sec:methods}
%%%%%%%%%%%%%%%%%%%%%%%%%%%%%%%%%%%%%%%%%%%%%%%%%%%%%%%%%%%%%%%%%%%%%%%%%%%%
\subsection{Implementation of the sparse linear operator as a constrained minimization in \texttt{LAMMPS}}
\label{sec:methods-sparse}
%%%%%%%%%%%%%%%%%%%%%%%%%%%%%%%%%%%%%%%%%%%%%%%%%%%%%%%%%%%%%%%%%%%%%%%%%%%%

Equation (\ref{lin_op}) defines a linear operator which can be used in iterative solution of the linear equations (\ref{eq:impl_der_matrix}) and 
thus to evaluate the implicit derivative $\nabla_\T \tX_\T^*$. To fully exploit
the efficiencies afforded by the Hessian sparsity, in this section, we detail 
how $\nabla_\T \tX_\T^*$ can be implemented in the massively parallel
\texttt{LAMMPS} simulation package. This is achieved through $N_D$ constrained 
minimizations, one for each row 
$[\nabla_\T \tX_\T^*]_l$, $l\in[1,N_D]$ of the implicit derivative. 
%We have established, through a wide range of numerical tests using the full dense solution, that the off-diagonal terms $\nabla^2_{\tX\epsilon}\oU$ in (\ref{eq:impl_der_matrix}) can be neglected 
%when determining the homogeneous term $\nabla_\T\epsilon^*_\T$, 
%meaning the inhomogeneous term $\nabla_{\Theta_l} \tX_\T^*$ satisfies
As follows from equations (\ref{eq:impl_der_matrix}), the inhomogeneous term $\nabla_{\Theta_l} \tX_\T^*$ satisfies
\begin{equation}
    [\nabla_\T \tX_\T^*]_l\nabla^2_{\tX\tX}\oU
    +
    [{\bf B}]_l = {\bf 0},
\end{equation}
where ${\bf B}=\nabla_\T\epsilon^*_\T\otimes\nabla^2_{\epsilon\tX}\oU+\nabla^2_{\T\tX}\oU\in\mathbb{R}^{N_D\times N\times3}$.
For the linear-in-descriptor \texttt{SNAP} potentials
used here, the $\nabla^2_{\epsilon\T}\oU$ term can be directly accessed as derivatives of descriptors~\cite{Thompson_snap_2015} using  
\texttt{fix sna/atom} and related commands in \texttt{LAMMPS}.
For the inverse design applications in this paper, which modified only the interaction parameters of a single solute atom without changes to the supercell strain, we have $\nabla_\T\epsilon^*_\T={\bf0}$ and thus ${\bf B}=\nabla^2_{\T\tX}\oU$.
With an initial parameter vector $\T$, we then define the modified energy function for a parameter index $l$
\begin{equation}
    %U_{\alpha,l}(\tX,\C;\T) \equiv 
    U(\tX,\C;\T) + \alpha [{\bf B}]_l\cdot[\tX-\tX^*_\T].
    \label{eq:constrained}
\end{equation}
This modified energy is simple to implement in \texttt{LAMMPS}
through the \texttt{fix addforce} function, with similar ease
of implementation in any molecular dynamics package.
In the limit $\alpha\to0$, the minimizer $\tX^*_{l,\alpha}$ of (\ref{eq:constrained})
gives the $l^{\rm th}$ vector $[\nabla_\T \tX^*_{\T}]_l$ of the implicit derivative $\nabla_\T \tX^*_{\T}$ through
\begin{equation}
    [\nabla_\T \tX^*_{\T}]_l = \big(\tX^*_{l,\alpha} - \tX^*_\T\big) / \alpha.
\end{equation}
Further details regarding hyperparameter scanning for suitable values of $\alpha$ and comparison against 
the full dense linear solution employing the Hessian 
matrix is provided in supplementary material. 

\subsection{Multi-specie ML potentials }
\label{sec:methods-multispec}
For systems with multiple atomic species $s_i\in[1,N_S]$, as investigated in sections~\ref{sec:vac_form}, \ref{sec:inverse-design}, we
simply assign a new specie-dependent parameter vector as in the original \texttt{SNAP} paper~\cite{Thompson_snap_2015}, giving a total energy 
\begin{equation}
    U(\tX,\C;\T) = \sum_i [\T]_{s_i}{\bf D}(\tX,\C,i) \equiv \T\cdot\D(\tX,\C),
    \label{eqn:snap-multi-species}
\end{equation}
where $\T\in\mathbb{R}^{N_S\times N_D}$ is the total parameter vector, $s_i\in[1,N_S]$
and $\D(\tX,\C)\in\mathbb{R}^{N_S\times N_D}$ is the total descriptor vector.
In typical usage practice, the descriptor functions have species-dependent hyperparameters which must also be tuned, but here we assume these are fixed.

%%%%%%%%%%%%%%%%%%%%%%%%%%%%%%%%%%%%%%%%%%%%%%%%%%%%%%%%%%%%%%%%%%%%%%%%%%%%
\subsection{\textit{Ab initio} QM-ML calculations}
\label{sec:methods-QM-ML}
%%%%%%%%%%%%%%%%%%%%%%%%%%%%%%%%%%%%%%%%%%%%%%%%%%%%%%%%%%%%%%%%%%%%%%%%%%%%

The \textit{ab initio} reference data for Be segregation 
to screw dislocations in W was calculated using 
QM-ML hybrid simulations~\cite{Grigorev_2020,grigorev2023calculation},
which couple \textit{ab initio} and machine learning potentials. Initial structures were obtained with \texttt{matscipy.dislocation} module~\cite{Grigorev2024}. The Be segregation calculation used the same approach 
as for He segregation reported in~\citep{grigorev2023calculation}.
\textit{Ab initio} forces were evaluated using \texttt{VASP}~\cite{Kresse1996,Perdew1996}
with 10 $\mathbf{k}$-points along the periodic line direction, with a cutoff energy of 500 eV
and a minimization force threshold of 0.01~eV/$\text{\AA}$.
The machine learning force field was a modified \texttt{SNAP/MILaDy}
potential from~\cite{goryaeva2021}.
The QM/ML coupling used a buffer radius of 10~$\text{\AA}$, resulting in a total of 246 \texttt{VASP} atoms, of which 168 were in the buffer. 
%We refer the reader to~\citep{grigorev2023calculation} for further details.

\section{Data Availability}
The implicit derivative implementations derived will be publicly available on GitHub following peer review. 

\vspace{0.2cm}
\section{Acknowledgements}
IM and TDS gratefully acknowledge support from an Emergence@INP grant from the CNRS.
TDS thanks the Institute for Pure and Applied Mathematics at the University of California, Los Angeles (supported by NSF grant DMS-1925919) for their hospitality. TDS and PG gratefully acknowledge support from ANR grants ANR-19-CE46-0006-1 and ANR-23-CE46-0006-1, IDRIS allocation A0120913455.

\section{Contributions}
TDS designed the research program and derived the initial theoretical results. IM implemented the sparse operator, designed the implicit loss minimizer, and ran all simulations. PG generated the dislocation structures and performed the QM/ML calculations. IM and TDS wrote the paper.

\section{Competing interests}

The authors declare no competing interests.

%%%%%%%%%%%%%%%%%%%%%%%%%%%%%%%%%%%%%%%%%%%%%%%%%%%%%%%%%%%%%%%%%%%%%%%%%%%%

%%%%%%%%%%%%%%%%%%%%%%%%%%%%%%%%%%%%%%%%%%%%%%%%%%%%%%%%%%%%%%%%%%%%%%%%%%%%

\begin{widetext}
\newpage
\begin{center}
\Large Supplemental Material for:\\
Exploring the parameter dependence of atomic minima with implicit differentiation\\
\end{center}

\setcounter{section}{0}
\renewcommand{\thesection}{\arabic{section}}
%%%%%%%%%%%%%%%%%%%%%%%%%%%%%%%%%%%%%%%%%%%%%%%%%%%%%%%%%%%%%%%%%%%%%%%%%%%%
\section{IMPLICIT DIFFERENTIATION}
\vspace{0.2cm}
%%%%%%%%%%%%%%%%%%%%%%%%%%%%%%%%%%%%%%%%%%%%%%%%%%%%%%%%%%%%%%%%%%%%%%%%%%%%
\subsection{General expression derivation}
%%%%%%%%%%%%%%%%%%%%%%%%%%%%%%%%%%%%%%%%%%%%%%%%%%%%%%%%%%%%%%%%%%%%%%%%%%%%
\vspace{0.2cm}

Here we consider $\X^*_\T$ as a stationary configuration that includes the scaled atomic coordinates $\tX^*_\T$ and supercell $\C^*_\T$. A stationary configuration is defined with a zero-force condition:

\begin{equation}
    \F(\X^*_\T; \T) = {\bf 0}.
\end{equation}
Under a parameter perturbation, $\T + \delta \T$, a new stationary configuration $\X^*_{\T+\delta \T}$ will satisfy:
\begin{equation}
    \F(\X^*_{\T+\delta \T}; \T+\delta \T) = {\bf 0}.
\end{equation}
Hence, under a parameter perturbation, the deferential of the force is zero as well:
\begin{equation}
    \label{eq:force-diff}
    \mathrm{d} \F(\X^*_{\T+\delta \T}; \T+\delta \T) = \delta \X^*_\T \nabla_\X \F(\X^*_\T; \T) + \delta \T \cdot \nabla_\T \F(\X^*_\T; \T) + \mathcal{O}(\delta\T^2)= {\bf 0}.
\end{equation}
We express the variation in coordinates using the implicit derivative definition $\delta \X^*_\T=\delta\T \nabla_\T \X^*_\T$ and obtain
\begin{equation}
    \delta \T \cdot \big[ \nabla_\T \X^*_\T  \nabla_\X\F(\X^*_\T; \T) + \nabla_\T \F(\X^*_\T; \T) \big] = {\bf 0} + \mathcal{O}(\delta\T^2).
\end{equation}
As this holds for any parameter variation $\delta\T$, the term inside the square brackets is zero. Finally, we will express the force through energy $\oU(\X^*_\T; \T) = -\nabla_\X \F(\X^*_\T; \T)$ and get
\begin{equation}
    \label{eq:impl-der-full}
    %\nabla_\T \X^*_\T  \nabla_\X\F(\X^*_\T; \T) = - \nabla_\T \F(\X^*_\T; \T)
    \nabla_\T \X^*_\T  \nabla^2_{\X\X}\oU(\X^*_\T; \T) = - \nabla_{\T\X} \oU(\X^*_\T; \T).
\end{equation}
Expressing atomic coordinates as $\X = \tX\C$ and splitting the position and mixed Hessians on the scaled coordinate, off-diagonal, and supercell blocks, $\big(\nabla^2_{\tX\tX}\oU\big), \big(\nabla^2_{\tX\epsilon}\oU\big), \big(\nabla^2_{\epsilon\epsilon}\oU\big)$, one gets the equation (5) from the main text.

\vspace{0.5cm}
%%%%%%%%%%%%%%%%%%%%%%%%%%%%%%%%%%%%%%%%%%%%%%%%%%%%%%%%%%%%%%%%%%%%%%%%%%%%
\subsection{\textit{Homogeneous} implicit derivative of strain}
%%%%%%%%%%%%%%%%%%%%%%%%%%%%%%%%%%%%%%%%%%%%%%%%%%%%%%%%%%%%%%%%%%%%%%%%%%%%
\vspace{0.2cm}

In this work, we split the full implicit derivative (equation~(\ref{eq:impl-der-full})) onto the \textit{inhomogeneous} and \textit{homogeneous} contributions. As explained in the main text (methods A), the \textit{inhomogeneous} implicit derivative $\nabla_\T \tX_\T^*$ can be computed in \texttt{LAMMPS} with constraint minimization, or alternatively, with automatic differentiation (AD), section IID from the main text. Here, we detail our finite difference implementation of the \textit{homogeneous} part of the implicit derivative, $\nabla_\T\epsilon^*_\T$. 

A stationary configuration of a system of volume $V^*_\T$ corresponds to zero pressure:
\begin{equation}
   P(V^*_\T) = - \frac{\partial \oU(\tX^*_\T, \C^*_\T; \T)}{\partial V} \Bigg|_{V^*_\T} = 0.
\end{equation}
For isotropic supercell variations, $\C^*_\T=\left[1+ \epsilon^*_\T\right]\C_0$, this equation can be reformulated in terms of strain:
\begin{equation}
    \label{eq:dU-dEps}
    \frac{\partial \oU(\tX, \epsilon; \T)}{\partial \epsilon} \Bigg|_{\tX^*_\T, \epsilon^*_\T} = 0.
\end{equation}
In this section, we neglect the \textit{inhomogeneous} contribution to the strain derivative and we will omit $\tX$ from the arguments of $\oU$ for clarity. For linear-in-descriptor potentials (e.g. equation (12) in the main text), equation~(\ref{eq:dU-dEps}) writes
\begin{equation}
    \T \cdot \frac{\partial \D(\epsilon)}{\partial \epsilon} \Bigg|_{\epsilon^*_\T} = 0.
\end{equation}
Note that the descriptor vector does not depend on potential parameters $\T$.
In analogy with the previous section (equation~(\ref{eq:force-diff})), we consider a system upon a parameter variation $\T+\delta \T$:
\begin{equation}
    \label{eq:pressure-perturbed}
    \frac{\partial U(\epsilon; \T + \delta \T)}{\partial \epsilon} 
    \Bigg|_{\epsilon^*_\T + \delta \epsilon^*_\T} = 0.
\end{equation}
Applying the linear-in-descriptor form of potential energy, we get:
\begin{equation}
    (\T+\delta \T) \cdot \frac{\partial \D(\epsilon)}{\partial \epsilon}
    \Bigg|_{\epsilon^*_\T + \delta \epsilon^*_\T} = 0.
\end{equation}
Neglecting the term proportional to $\delta\T \delta \epsilon^*_\T$, we get
\begin{equation}
    \delta \T \cdot \frac{\partial \D(\epsilon)}{\partial \epsilon} \Bigg|_{\epsilon^*_\T}
    + \delta \epsilon^*_\T \T \cdot  \frac{\partial^2 \D(\epsilon)}{\partial \epsilon^2} \Bigg|_{\epsilon^*_\T} = 0.
\end{equation}
We then use the definition of the \textit{homogeneous} implicit derivative, $\delta \epsilon^*_\T = \delta \T \nabla_\T \epsilon_\T^*$:
\begin{equation}
    \delta \T \cdot \Bigg[ \frac{\partial \D(\epsilon)}{\partial \epsilon} 
    + \nabla_\T \epsilon_\T^* \Big( \T \cdot  \frac{\partial^2 \D(\epsilon)}{\partial \epsilon^2} \Big) \Bigg]_{\epsilon^*_\T} = 0.
\end{equation}
Since this equation is valid for any parameter variation $\delta \T$, we can get the final expression for the \textit{homogeneous} implicit derivative:
\begin{equation}
    \nabla_\T \epsilon_\T^* = - \frac{ \partial \D(\epsilon) / \partial \epsilon }{ \T \cdot \partial^2 \D(\epsilon) / \partial \epsilon^2} \Bigg|_{\epsilon^*_\T}.
\end{equation}
For numerical purposes, we evaluate the derivatives of the descriptor vector with finite differences: 

\begin{equation}
    \frac{\partial \D(\epsilon^*_\T)}{\partial \epsilon} \approx \frac{\D(\epsilon^*_\T + \Delta \epsilon)-\D(\epsilon^*_\T - \Delta \epsilon)}{2\Delta \epsilon}; \quad 
    \frac{\partial^2 \D(\epsilon^*_\T)}{\partial \epsilon^2} \approx \frac{\D(\epsilon^*_\T + \Delta \epsilon)+\D(\epsilon^*_\T - \Delta \epsilon)-2\D(\epsilon^*_\T)}{\Delta \epsilon^2},
\end{equation}
where $\Delta \epsilon$ is typically $10^{-3} \text{\AA}$.

\vspace{0.5cm}
%%%%%%%%%%%%%%%%%%%%%%%%%%%%%%%%%%%%%%%%%%%%%%%%%%%%%%%%%%%%%%%%%%%%%%%%%%%%
\subsection{Implicit derivative of strain including \textit{homogeneous} and \textit{inhomogeneous} terms}
%%%%%%%%%%%%%%%%%%%%%%%%%%%%%%%%%%%%%%%%%%%%%%%%%%%%%%%%%%%%%%%%%%%%%%%%%%%%
\vspace{0.2cm}

Let us take into account the \textit{inhomogeneous} contribution in equation~(\ref{eq:pressure-perturbed}):
\begin{equation}
    \label{eq:pressure-perturbed-implicit}
    \frac{\partial U(\tX^*_\T+\delta\tX^*_\T, \epsilon^*_\T + \delta \epsilon^*_\T; \T + \delta \T)}{\partial \epsilon} = 0.
\end{equation}
Following a similar procedure as in the previous section, we get the \textit{h+ih} level of approximation for the strain implicit derivative:
\begin{equation}
    \nabla_\T \epsilon_\T^* = - \frac{ 
    \partial \D(\epsilon) / \partial \epsilon 
    + \nabla_\T \X^*_\T\, \partial \F / \partial \epsilon
    }
    { \T \cdot \partial^2 \D(\epsilon) / \partial \epsilon^2}\Bigg|_{\tX^*_\T,\epsilon^*_\T},
\end{equation}
where the force derivative over the strain is evaluated with the finite difference approach.

\vspace{0.5cm}
%%%%%%%%%%%%%%%%%%%%%%%%%%%%%%%%%%%%%%%%%%%%%%%%%%%%%%%%%%%%%%%%%%%%%%%%%%%%
\subsection{Taylor expansion of energy}
%%%%%%%%%%%%%%%%%%%%%%%%%%%%%%%%%%%%%%%%%%%%%%%%%%%%%%%%%%%%%%%%%%%%%%%%%%%%
\vspace{0.2cm}

In this section, we derive the expansion of potential energy of a stationary system, $\oU(\X^*_\T; \T)$, under a potential parameter perturbation. For clarity, we will omit $(\X^*_\T; \T)$ from $\oU(\X^*_\T; \T)$ in this section. For the expansion, one has to account for contributions arising from the explicit dependence of $\oU$ on parameters $\T$, but also from the change in stationary configuration $\X^*_\T$. Here, we first consider the general atomic coordinates $\X^*_\T$ and later split them on scaled coordinates $\tX^*_\T$ and supercell $\C^*_\T$.

Under a parameter perturbation $\T+\delta \T$, the potential energy expansion reads 

\begin{equation}
\begin{split}
    \oU(\X^*_\T + \delta \X^*_\T; & \T + \delta \T) = 
    \oU
    + \delta \X^*_\T \nabla_\X \oU + \delta \T \nabla_\T \oU \\
   &+ \frac{1}{2} \delta \X^*_\T \nabla^2_{\X\X} \oU {\delta \X^*_\T}^{\top}
    + \frac{1}{2}  \delta \T \nabla^2_{\T\T} \oU \delta \T^{\top}
    + \delta \T \nabla^2_{\T\X} \oU {\delta \X^*_\T}^{\top}
    + \mathcal{O}(\delta \T^3)
\end{split}
\end{equation}
The term proportional to $\nabla_\X \oU$ vanishes since $\X^*_\T$ is a stationary atomic configuration. For the second-order terms, we express $\delta \X^*_\T$ using the implicit derivative: $\delta \X^*_\T = \delta \T \nabla_\T \X^*_\T$. We further use the implicit derivative expression, equation~(\ref{eq:impl-der-full}), and get: $\delta \X^*_\T =-\delta \T \nabla^2_{\T\X} \oU \big[\nabla^2_{\X\X} \oU\big]^+$. The final energy expansion up to terms $\mathcal{O}(\delta \T^3)$ reads:

\begin{equation}
\begin{split}
     \oU(\X^*_\T &+ \delta \X^*_\T; \T + \delta \T) = \oU + \delta \T \nabla_\T \oU 
    + \frac{1}{2} \delta \T \Big[ \nabla^2_{\T\T} \oU + \nabla^2_{\T\X} \oU (\nabla_\T \X^*_\T)^{\top} \Big] \delta \T^{\top}.   
\end{split}
\end{equation}

In the main text, we outline the energy expansion as follows:
\begin{equation}
    \delta^{(\zeta)}\oU^* 
    \equiv\delta \T \nabla_{\T} \oU
    + \delta \T {\bf H}_{\zeta} \delta \T^\top + \mathcal{O}(\delta\T^3),
    \label{eq:energy}
\end{equation}
where $\zeta$ represents the level of approximation. Considering $\X = \tX\C$ and employing the implicit derivative definitions for scaled coordinates and strain, $\nabla_\T \tX_\T^*$ and $\nabla_\T \epsilon_\T^*$ (equations (2)-(5) from the main text), we derive the expressions of ${\bf H}_{\zeta}$ corresponding to each level of approximation:

\begin{itemize}
    \item \textit{constant (c)}: ${\bf H}_{c} = \nabla^2_{\T\T} \oU$
    \item \textit{homogeneous (h)}: ${\bf H}_{h} = \nabla^2_{\T\T} \oU + \nabla^2_{\T\epsilon} \oU (\nabla_\T \epsilon^*_\T)^{\top}$
    \item \textit{inhomogeneous (ih)}: ${\bf H}_{ih} = \nabla^2_{\T\T} \oU + \nabla^2_{\T\tX} \oU (\nabla_\T \tX^*_\T)^{\top}$.
\end{itemize}

\vspace{0.5cm}
%%%%%%%%%%%%%%%%%%%%%%%%%%%%%%%%%%%%%%%%%%%%%%%%%%%%%%%%%%%%%%%%%%%%%%%%%%%%
\section{EFFICIENCY OF IMPLICIT DERIVATIVE EVALUATION SCHEMES}
%%%%%%%%%%%%%%%%%%%%%%%%%%%%%%%%%%%%%%%%%%%%%%%%%%%%%%%%%%%%%%%%%%%%%%%%%%%%
\vspace{0.2cm}

In this section, we provide the details on memory and time efficiency of the automatic differentiation (AD) and \texttt{LAMMPS} implementations of the implicit derivative. 
Given its computational complexity, our focus here will be on the \textit{inhomogeneous} contribution, $\nabla_\T \tX_\T^*$.

\vspace{0.5cm}
%%%%%%%%%%%%%%%%%%%%%%%%%%%%%%%%%%%%%%%%%%%%%%%%%%%%%%%%%%%%%%%%%%%%%%%%%%%%
\subsection{Automatic differentiation implementation}
%%%%%%%%%%%%%%%%%%%%%%%%%%%%%%%%%%%%%%%%%%%%%%%%%%%%%%%%%%%%%%%%%%%%%%%%%%%%
\vspace{0.2cm}

Here, we use the LJ random fcc alloy (presented in the main text, section IIF) as a test system.
We compute the implicit derivative with AD using three approaches: 1) Computing the pseudo-inverse of the Hessian matrix with dense linear algebra solution, called \textit{dense}. 2) Finding $\X^*_{\T+\delta \T}$ with minimization (gradient descent method was used in this work) and applying AD to the entire pipeline of functions (potential energy and its derivatives) with \texttt{jaxopt} Python library,
%~\cite{ablin2020super,blondel2022efficient}, 
called \textit{jaxopt}. 3) Sparse linear operator technique (section IIC in the main text) within the AD framework, called \textit{sparse}.

\begin{figure}[h!]
    \centering
    \includegraphics[width=0.8\columnwidth]{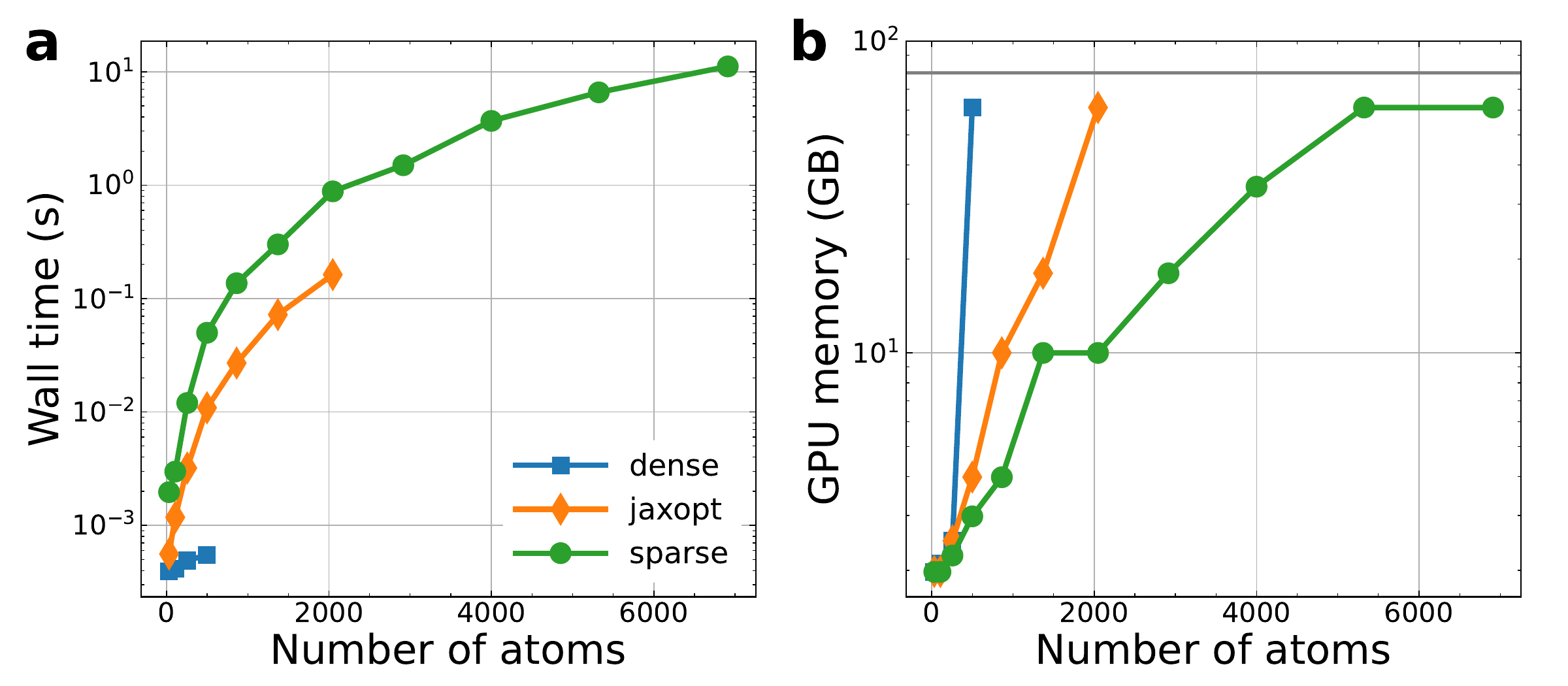}
    \caption{
    \textbf{Efficiency of AD implicit derivative solvers.}
    \textbf{a} Wall time and and \textbf{b} GPU memory required to compute the \textit{inhomogeneous} implicit derivative vs the number of atoms in an LJ random alloy system.
    The gray horizontal line on panel \textbf{b} indicates the GPU memory limit of NVIDIA A100 80 GB graphic card.}
    \label{fig:autodiff-profiling}
\end{figure}

For this study, we used the best-in-class NVIDIA A100 80 GB graphic card tailored for scientific computations. To track the usage of the GPU memory for each solver and system size, we used the NVIDIA Management Library, \texttt{nvml} through the Python interface provided by the \texttt{pynvml} package.
Figure~\ref{fig:autodiff-profiling} shows the time (panel~\textbf{a}) and memory (panel~\textbf{b}) required for \textit{inhomogeneous} implicit derivative evaluation as a function of number of atoms in the system. The \textit{inverse} approach shows the best time performance, however, it runs out of 80 GB GPU memory at a system of 500 atoms. The second fastest, \textit{jaxopt} method, allows one to achieve system sizes of up to 2000 atoms before saturating the memory. Lastly, \textit{sparse} technique, is the most memory efficient and reaches the systems of up to 7000 atoms on a single A100 GPU. 

We would like to emphasize that while the methods based on AD provide unique advantages such as computational efficiency and ease of implementation, their substantial memory consumption significantly limits their applicability for large-scale simulations.

\vspace{0.5cm}
%%%%%%%%%%%%%%%%%%%%%%%%%%%%%%%%%%%%%%%%%%%%%%%%%%%%%%%%%%%%%%%%%%%%%%%%%%%%
\subsection{Implicit derivative implementation using LAMMPS package}
%%%%%%%%%%%%%%%%%%%%%%%%%%%%%%%%%%%%%%%%%%%%%%%%%%%%%%%%%%%%%%%%%%%%%%%%%%%%
\vspace{0.2cm}

Here, we discuss the time and memory efficiency of the \textit{inhomogeneous} implicit derivative implementation using the \texttt{LAMMPS} software. The test system is a vacancy in bcc tungsten with \texttt{SNAP} potential (section IIG in the main text).
We used four CPU nodes with two MD EPYC 7763 CPUs and 512 GB of RAM per node. We monitored RAM usage with the \texttt{vmstat} tool. We test the efficiency of the \textit{dense} and \textit{sparse} approaches presented above for AD. Additionally, we explore the efficiency of the constraint energy minimization approach, called \textit{energy}, that is described in the main text (section IVA). As seen from Figure~\ref{fig:LAMMPS-profiling}, the energy approach is the most time- and memory-efficient for large systems. Due to the efficient massive parallelization of the \texttt{LAMMPS} software, this method can be effectively scaled, allowing implicit derivative evaluation for systems of arbitrary size.

\begin{figure}[h!]
\centering
\includegraphics[width=0.8\columnwidth]{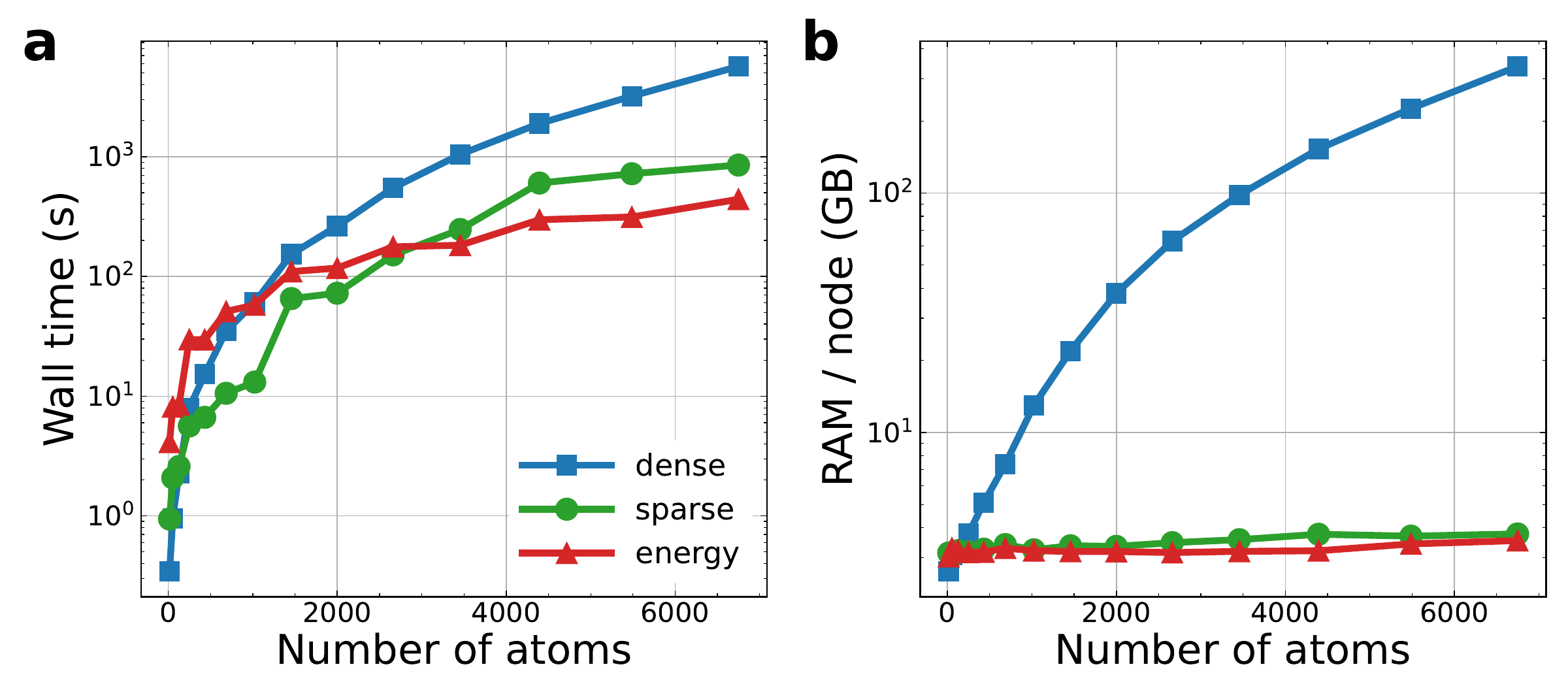}
\caption{
\textbf{Efficiency of LAMMPS implicit derivative implementation approaches.}
\textbf{a} Wall time and \textbf{b} RAM required to compute the \textit{inhomogeneous} implicit derivative as a function of the number of atoms in a vacancy in a bcc tungsten system with \texttt{SNAP} potential.}  
\label{fig:LAMMPS-profiling}
\end{figure}

\vspace{0.5cm}
%%%%%%%%%%%%%%%%%%%%%%%%%%%%%%%%%%%%%%%%%%%%%%%%%%%%%%%%%%%%%%%%%%%%%%%%%%%%
\section{LAMMPS IMPLEMENTATION: ACCURACY AND NUMERICAL DETAILS}
%%%%%%%%%%%%%%%%%%%%%%%%%%%%%%%%%%%%%%%%%%%%%%%%%%%%%%%%%%%%%%%%%%%%%%%%%%%%
\vspace{0.2cm}

%%%%%%%%%%%%%%%%%%%%%%%%%%%%%%%%%%%%%%%%%%%%%%%%%%%%%%%%%%%%%%%%%%%%%%%%%%%%
\subsection{Accuracy of position change predictions}
%%%%%%%%%%%%%%%%%%%%%%%%%%%%%%%%%%%%%%%%%%%%%%%%%%%%%%%%%%%%%%%%%%%%%%%%%%%%
\vspace{0.2cm}

In this section, we present a comparison of three methods for evaluating the \textit{inhomogeneous} implicit derivative, \textit{dense}, \textit{sparse}, and \textit{energy}, as discussed above, implemented within the \texttt{LAMMPS} package. The test system is a vacancy in bcc tungsten and \texttt{SNAP} potential parameters are set according to the section IIG of the main text, equation~(13), $\delta \T=\lambda(\T_m - \bar{\T})$, with a strong perturbation of $\lambda=40$ and one selected potential sample $m$. The predicted position changes are computed as
\begin{equation}
    \delta \tX^{*\;\rm pred} = \tX^* + \delta \T \nabla_\T \tX^*_\T,
\end{equation}
where the \textit{inhomogeneous} implicit derivative $\nabla_\T \tX^*_\T$ is computed with one of the three method and positions $\tX^*$ are obtained through the energy minimization of the system with the reference potential $\bar{\T}$. The true position changes are calculated as follows
\begin{equation}
    \delta \tX^{*\;\rm true} = \tX^*(\T) - \tX^*(\bar{\T}),
\end{equation}
where $\tX^*(\T)$ are the minimized positions of a system with parameters $\T$ using \texttt{LAMMPS}, the minimum image convention is applied to coordinates.

As shown in Fig.~\ref{fig:impl-der-numeric}a, there is a remarkable agreement between the true and predicted positions, with negligible differences among the three computational methods. Given the low computational cost and memory requirements, the \textit{energy} method stands out as the optimal choice for \textit{inhomogeneous} implicit derivative evaluation.

\begin{figure}[h!]
    \centering
    \includegraphics[width=0.8\columnwidth]{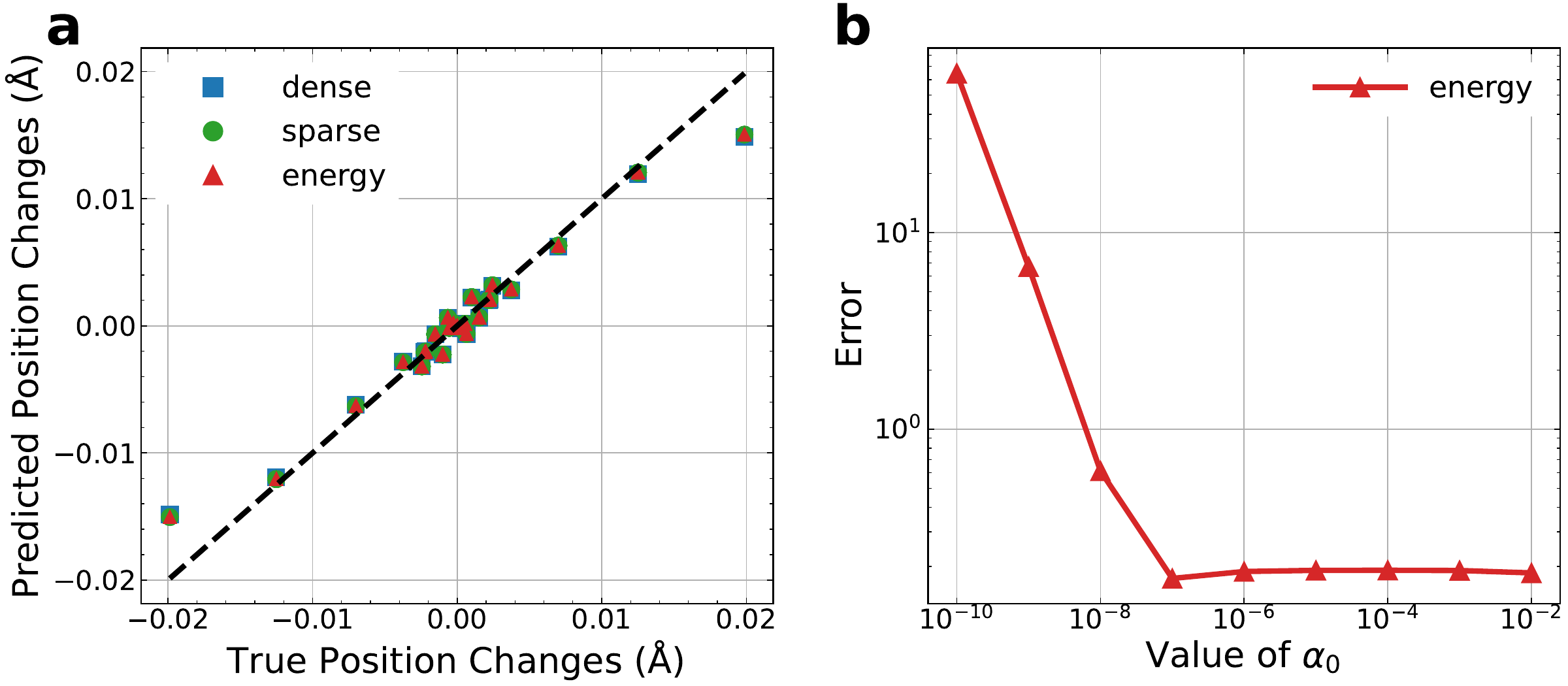}
    \caption{ 
    \textbf{Accuracy of the \textit{inhomogeneous} implicit derivative solvers implemented in \texttt{LAMMPS}.}
    \textbf{a} Predicted vs true position changes for the vacancy in the bcc tungsten system.
    \textbf{b} Error of the \textit{inhomogeneous} position changes prediction for the \textit{energy} method as a function of the $\alpha_0$ parameter.
    }
    \label{fig:impl-der-numeric}
\end{figure}

\subsection{Numerical details of LAMMPS constraint energy minimization}
\vspace{0.2cm}

Here, we detail our constraint energy minimization approach (called \textit{energy}) to compute the \textit{inhomogeneous} implicit derivative implemented in \texttt{LAMMPS}. As explained in the main text, the implicit derivative corresponding to a parameter index $l$ is obtained as $[\nabla_\T \tX^*_{\T}]_l = \big(\tX^*_{l,\alpha} - \tX^*_\T\big) / \alpha$. We have found that the optimal values of  $\alpha$ should be computed for each parameter $\T_l$ separately as follows
\begin{equation}
    \alpha(l) = \frac{\alpha_0}{\max{\big(\big| \big[ \nabla^2_{\T\tX}\oU\big]_l \big|\big)} },
\end{equation}
where $\alpha_0$ is a constant and $\big[ \nabla^2_{\T\tX}\oU\big]_l$ is a row $l$ of the mixed Hessian.

Figure~\ref{fig:impl-der-numeric}b presents the dependence of the error of the implicit derivative prediction as a function of the $\alpha_0$ parameter. The error is computed as
\begin{equation}
    \frac{\| \delta \tX^{*\;\rm true} - \delta \tX^{*\;\rm pred}  \| }{\| \delta \tX^{*\;\rm true}\|}.
\end{equation}
For $\alpha_0<10^{-7}$, the error is large due to limitations in machine precision. For larger $\alpha_0$ values, the error does not change. However, the number of energy minimization iterations increases significantly for $\alpha_0\ge10^{-3}$. Therefore, we conclude that values of $\alpha_0\in[10^{-6};10^{-4}]$ are optimal for the \textit{energy} implicit derivative method. For the constraint energy minimization, we use the \texttt{FIRE} algorithm implemented in \texttt{LAMMPS}.

\vspace{0.5cm}
%%%%%%%%%%%%%%%%%%%%%%%%%%%%%%%%%%%%%%%%%%%%%%%%%%%%%%%%%%%%%%%%%%%%%%%%%%%%
\section{INVERSE DESIGN}
%%%%%%%%%%%%%%%%%%%%%%%%%%%%%%%%%%%%%%%%%%%%%%%%%%%%%%%%%%%%%%%%%%%%%%%%%%%%
\vspace{0.2cm}
\subsection{Adaptive step for loss minimization}
%%%%%%%%%%%%%%%%%%%%%%%%%%%%%%%%%%%%%%%%%%%%%%%%%%%%%%%%%%%%%%%%%%%%%%%%%%%%
\vspace{0.2cm}

This section describes the adaptive step calculation for the inverse design applications. As explained in the main text (section IIIH), at iteration $k+1$ of the minimization procedure, the potential parameters are updated as $\T^{(k+1)} = \T^{(k)} - h \nabla_{\T} L(\T^{(k)})$ and positions as $\X^*_{\T^{(k+1)}} = \X^*_{\T^{(k)}} + (\T^{(k+1)} - \T^{(k)}) \nabla_{\T} \X^{*}_{\T^{(k)}} =  \X^*_{\T^{(k)}} - h \nabla_{\T} L(\T^{(k)}) \nabla_{\T} \X^{*}_{\T^{(k)}}$. Accordingly, the loss at iteration $k+1$ is
\begin{equation}
    L(\T^{(k+1)}) = \frac{1}{2} \| \X^*_{\T^{(k)}} + h\Delta \X^{(0)}_{\T^{(k)}} - \X^*_{\rm ref} \|^2,
\end{equation}
where $\Delta \X^{(0)}_{\T^{(k)}}$ is defined as the change in atomic positions corresponding to step $h=1$, $\Delta \X^{(0)}_{\T^{(k)}} \equiv -\nabla_{\T} L(\T^{(k)}) \nabla_{\T} \X^{*}_{\T^{(k)}}$. Then, the \textit{change} in loss at a given iteration is
\begin{equation}
    \Delta L(\T^{(k+1)}) \equiv  L(\T^{(k+1)}) -  L(\T^{(k)}) = h \Delta \X^{(0)\,\top}_{\T^{(k)}} \big(\X^*_{\T^{(k)}} - \X^*_{\rm ref}\big) + \frac{1}{2} h^2 
    \| \Delta \X^{(0)}_{\T^{(k)}}\|^2.
\end{equation}
Finally, the step $h(k)$ that minimizes the loss at iteration $k$ can be found as
\begin{equation}
    h(k) = - \frac{ \Delta \X^{(0)\,\top}_{\T^{(k)}} \big(\X^*_{\T^{(k)}} - \X^*_{\rm ref}\big) }
    { \| \Delta \X^{(0)}_{\T^{(k)}}\|^2 }.
\end{equation}
Accordingly, the atomic position change at iteration $k+1$ is 
\begin{equation}
    \Delta \X_{\T^{(k)}} = h(k) \nabla_\T L(\T^{(k)}) \nabla_\T \X^*_{\T^{(k)}}.
\end{equation}

\vspace{0.5cm}
%%%%%%%%%%%%%%%%%%%%%%%%%%%%%%%%%%%%%%%%%%%%%%%%%%%%%%%%%%%%%%%%%%%%%%%%%%%%
\subsection{W-Be Potential fine-tuning}
%%%%%%%%%%%%%%%%%%%%%%%%%%%%%%%%%%%%%%%%%%%%%%%%%%%%%%%%%%%%%%%%%%%%%%%%%%%%
\vspace{0.2cm}

Figure~\ref{fig:WBe-minim-stats} presents the loss minimization for the potential fine-tuning for the W-Be system presented in the main text (section IIH).

\begin{figure}[h!]
    \centering
    \includegraphics[width=0.8\columnwidth]{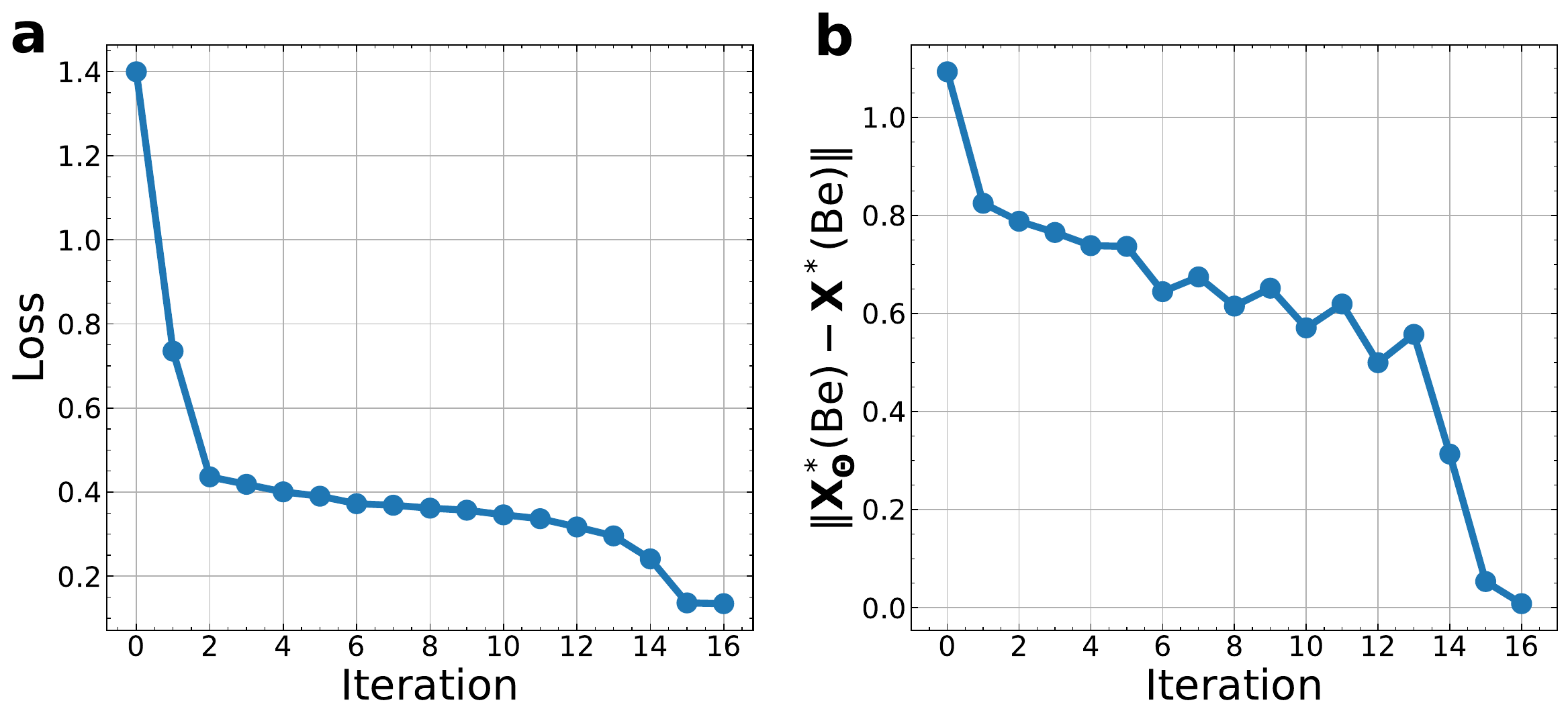}
    \caption{ 
    \textbf{Loss minimization for W-Be potential fine-tuning.} 
    \textbf{a} Implicit loss during the minimization procedure as defined in equation (9) of the main text.
    \textbf{b} Difference between a current and target Be atom positions during minimization.
    }
    \label{fig:WBe-minim-stats}
\end{figure}

\end{widetext}

\end{document}